\documentclass[aps,twocolumn,superscriptaddress,nofootinbib]{revtex4}

\usepackage{amsfonts}
\usepackage{amssymb}
\usepackage{mathdots}

\usepackage{graphicx}
\usepackage{bbm}
\usepackage{color}
\usepackage{pifont}
\usepackage{enumerate}
\usepackage{subfigure}
\usepackage{color}
\newcommand*{\rom}[1]{\expandafter\@slowromancap\romannumeral #1@}
\makeatother
\usepackage{mathptmx}
\usepackage{mathrsfs}
\usepackage{mathtools}
\usepackage{amsmath}
\usepackage{amsfonts}
\usepackage{amssymb}
\usepackage{mathdots}
\usepackage{bbold}
\usepackage{isomath}
\usepackage{mathrsfs}
\usepackage{amsthm} 
\usepackage{graphicx}
\usepackage{bbm}
\usepackage{color}
\usepackage{pifont}
\usepackage{enumerate}
\usepackage{amsthm}
\theoremstyle{plain}
\theoremstyle{definition}
\usepackage{mathtools}
\makeatletter
\newsavebox{\@brx}
\newcommand{\llangle}[1][]{\savebox{\@brx}{\(\m@th{#1\langle}\)}%
  \mathopen{\copy\@brx\kern-0.5\wd\@brx\usebox{\@brx}}}
\newcommand{\rrangle}[1][]{\savebox{\@brx}{\(\m@th{#1\rangle}\)}%
  \mathclose{\copy\@brx\kern-0.5\wd\@brx\usebox{\@brx}}}
\makeatother
\usepackage[titletoc]{appendix}
\usepackage[english]{babel}
\usepackage[autostyle]{csquotes}
\usepackage{hyperref} 
\usepackage{cleveref} 

\begin{document}

\title{Dipolar Relaxation Mechanism of Long Lived States in Methyl Groups}


\author{R. Annabestani}
\email[]{rannabes@uwaterloo.ca}
\affiliation{Institute for Quantum Computing, University of Waterloo, Waterloo, Ontario N2L 3G1, Canada}
\affiliation{Department of Physics and Astronomy, University of Waterloo, Waterloo, Ontario N2L 3G1, Canada}
\author{D. G. Cory}
\email[]{dcory@uwaterloo.ca}
\affiliation{Institute for Quantum Computing, University of Waterloo, Waterloo, Ontario N2L 3G1, Canada}
\affiliation{Department of Physics and Astronomy, University of Waterloo, Waterloo, Ontario N2L 3G1, Canada}
\affiliation{Department of Chemistry, University of Waterloo, Waterloo, Ontario N2L 3G1, Canada}
\affiliation{Perimeter Institute for Theoretical Physics, Waterloo, Ontario N2L 2Y5, Canada}
\affiliation{Canadian Institute for Advanced Research, Toronto, Ontario  M5G 1Z8, Canada}



\begin{abstract}
 We analyze the symmetry properties of the dipolar Hamiltonian as the main relaxation mechanism responsible for the observed NMR spectra of long lived states in methyl groups. Long lived states exhibit relaxation times that are considerably longer than the spin-lattice relaxation time, $T_{1}$. The analysis brings clarity to the key components of the relaxation mechanism of long lived states by focusing exclusively on the symmetry of the spin Hamiltonian. Our study shows that the dipole-dipole coupling between protons of a methyl group and between the protons and an external spin are both symmetry breaking interactions that can lead to relaxation pathways that transform the polarization from symmetry order to Zeeman order, but the net contribution of the internal dipolar interaction to the NMR observation of long lived states is zero. Our calculation is in agreement with the reported features of the observed spectra.  
 \end{abstract}


\maketitle
\section{Introduction}
Implementing states with long relaxation times is an attractive approach to quantum memories, as sensors in quantum meteorology and as a tool for studies of slow processes in nuclear magnetic resonance (NMR) \cite{CDAB05, Detal05, SVB07, ASVB09}. \\

NMR is normally concerned with states polarized along the external magnetic field at some temperature, and hence, the spin-lattice interaction is the dominant source of energy relaxation with its characteristic time $T_{1}$. There are, however, alternative ways of polarizing the spin system such that the polarized state is immune to the majority of relaxation mechanisms and the relaxation time can substantially exceed $T_{1}$. For instance, a group of indistinguishable spins have symmetry properties that are preserved under collective noise processes. These symmetry degrees of freedom can be used for implementing long lasting states \cite{L14}. \\

Long lived states (LLS) have been implemented experimentally \cite{CL04, PCHL08, Aetal09, C14, Feng14, M13}. For the simplest case of two spin half particles, the four spin eigenstates can be categorized into two symmetry groups of triplet states and singlet state, where the former is invariant under spin exchange and the later is antisymmetric. By creating an imbalance of population between the singlet state and the identity in the triplet subspace, one can prepare a \textit{symmetry polarized} state that exhibits long relaxation time due to its immunity to the majority of relaxation mechanisms \cite{CL04, CJL04}. One can extend this idea to three identical spins, where the eight eigenstates are categorized into three groups labelled $A$, $E_{+}$ and $E_{-}$ where the last two are degenerate subspaces. Under cyclic permutation of the spins, the four eigenstates in group $A$ remain invariance, and hence, are \textit{totally symmetric} states; whereas, the two eigenstates in each of degenerate subspaces, $E_{\pm}$, acquire a phase $\epsilon= e^{\pm \frac{2 \pi}{3}}$, and hence, are \textit{non-symmetric} states \cite{C71, IB12, M13}. In a similar manner to the two spin case, one can polarize three identical spins with respect to their symmetry by creating an imbalance of population between the $A$ states and the $E_{\pm}$ states. This has been implemented in methyl groups and the relaxation time has been extended by factor of 7 beyond the single spin $T_{1}$ \cite{ M13}.\\

The immunity to noise stems from the fact that the symmetry of states is preserved under collective noise operators. In order to \textit{prepare} a LLS one needs to make spins indistinguishable, for example, by means of themal cooling or by applying radio frequency pulses \cite{CL04, M13}. It is important to note that the entire space of a LLS need not be protected against noise. In fact, although the symmetry is preserved under collective noise processes, the magnetization is not. Therefore, in principle, a long lived state that is fully polarized in terms of its symmetry can be fully unpolarized in terms of its collective magnetization. As a result, a LLS is not necessarily an observable in NMR. However, one can \textit{observe} a LLS if the spins become distinguishable. In fact, any interaction that does not preserve the permutational symmetry of spins, in principle, can lead to a relaxation process that turns the polarization from \textit{symmetry order} into \textit{Zeeman order}, and hence, makes an initially prepared LLS an NMR observable. For example, in \cite{M13}, accessing the noise protected subspaces has been achieved at low temperature where the majority of those interactions that distinguishes spins are \textit{frozen} or have negligible effect, but, the NMR observation has been done at room temperature where those spin distinguishing relaxation processes are no longer negligible. A few examples of such interactions are the chemical shift anisotropy (CSA) and the dipole-dipole coupling between spins. The later is the subject of our study, where we analyze the symmetry properties of the dipolar Hamiltonian of methyl groups and study the contribution of this interaction to the NMR observation of long lived states. \\

The spectrum of initially prepared LLS in methyl groups has been reported and has some unique features \cite{IB12, M13} that differentiates an initial symmetry polarized state from a thermally polarized state. The scalar coupling between the methyl group with total spin operator $\vec{\textbf{S}}= \frac{1}{2} \sum\limits_{i=1}^{3}\ \vec{\sigma}^{H}_{i}$ and the carbon with spin operator $\vec{I}= \frac{1}{2} \sigma^{C}$ is given by $ 2\pi\ J_{HC}  \ \vec{\textbf{S}}\ .\ \vec{I}$, which leads to splitting the proton and the carbon spectra into multiple peaks. On the proton channel one would observe two resolved peaks, $\langle S_{z} \otimes E_{\uparrow}\rangle$ and $\langle S_{z} \otimes E_{\downarrow}\rangle$, each corresponding to the carbon's state being aligned or anti-aligned with the field, and, on the carbon channel, one would observe four resolved peaks, $\langle \Pi_{\pm\frac{ 3}{2}} \otimes I_{z}\rangle$ and $\langle \Pi_{\pm\frac{ 1}{2}} \otimes I_{z}\rangle$, each corresponding to the total magnetization of the methyl group being $\pm \frac{3}{2}$ or $\pm\frac{1}{2}$. In case of symmetry polarized states, the two peaks on the proton channel have equal amplitudes but with opposite phases and the four peaks on the carbon channel have different amplitudes and they are two by two anti-phase with one another \cite{LSMG10,M13}, as opposed to thermally polarized states where all peaks on both spectra are in-phase with each other.  In this report, we analytically calculate the NMR spectra of an initially prepared LLS and predict the above described features that have been observed. The authors of \cite{JLevitt15} have also studied the theory of preparation and observation of LLS in methyl groups. Our approach relies on the symmetry properties of the dipolar Hamiltonian and the noise protected states and can be considered as complementary to previous studies \cite{M13, JLevitt15}. \\


In section \ref{Sec_LLS}, we provide a definition for long lived states in the case of three indistinguishable spins. In section \ref{Sec_Dipolar} we expand both the hetronuclear and homonuclear dipolar Hamiltonians in terms of symmetrized spin operators and analyze their symmetry properties. The symmetry analysis allows us to solve the master equation analytically in section \ref{Sec_MasterEq} and to predict the NMR spectrum of LLS in methyl groups. Finally, we report a summary of this work in section \ref{Sec_Conclude}.
\section{Long Lived State in Methyl Groups}
\label{Sec_LLS}
In the presence of a uniform static magnetic field, $B_{0}$, the spin Hamiltonian of a methyl group is
 \begin{equation}
 \label{Eq_Spin_Ham}
 H_{\text{spin}}= \frac{\omega_{h}}{2} \sum\limits_{i=1}^{3} \ \sigma^{(i)}_{z} + 2\pi \
 J_{0}\ \sum\limits_{j< k} \  \vec{\sigma}^{(j)}.\vec{\sigma}^{(k)} + H_{CSA} + H_{DD},
 \end{equation}
 where $\omega_{h}= \gamma_{h}B_{0}$ is the proton frequency, $J_{0}$ is the scalar coupling constant between any two protons and $H_{CSA}$ and $H_{DD}$ account for the chemical shift anisotropy (CSA) and the dipole-dipole (DD) interaction between any two spins. At relatively large field, when the Zeeman interaction is the dominant term in the Hamiltonian, it is a good approximation to treat these protons as three indistinguishable spins. The spin eigenstates are
 
 \begin{eqnarray}
  |A, 3/2\rangle &=&|\uparrow \uparrow \uparrow\rangle \\ \nonumber
|A, 1/2\rangle &=&\frac{1}{\sqrt{3}}\left(| \uparrow\uparrow\downarrow \rangle+ | \downarrow \uparrow \uparrow\rangle + |\uparrow \downarrow\uparrow  \rangle \right) \\\nonumber
|E_{+}, 1/2\rangle &=&\frac{1}{\sqrt{3}}\left(| \uparrow\uparrow\downarrow\rangle+ \epsilon^{*}|  \downarrow \uparrow \uparrow \rangle +  \epsilon |\uparrow \downarrow\uparrow  \rangle \right) \\\nonumber
|E_{-}, 1/2\rangle &=&\frac{1}{\sqrt{3}}\left(| \uparrow\uparrow\downarrow\rangle+ \epsilon|  \downarrow \uparrow \uparrow \rangle +  \epsilon ^{*}|\uparrow \downarrow\uparrow  \rangle \right) \\\nonumber
 \end{eqnarray}
with $\epsilon=e^{i \frac{2\pi}{3}}$. The other four eigenstates are obtained by replacing $|\uparrow\rangle  \leftrightarrow |\downarrow\rangle$. We denote the spin eigenstates by $\{|s, m\rangle\}$ where the first label $s\in \{A, E_{+}, E_{-}\}$  is the "symmetry label" and the second label  $m \in \{\pm \frac{3}{2}, \pm \frac{1}{2} \}$ is the collective magnetization. The symmetry label reflects that if we cyclically permute the indistinguishable spins, the $s=A$ eigenstates are invariant and the $s= E_{+}$ ($s=E_{-}$) states acquire a phase $\epsilon$ (or $\epsilon^{*}$). In addition, all $|s, m\rangle$ are eigenstates of the Zeeman Hamiltonian, and hence, they are labelled by the eigenvalues of the $z$ component of the total spin angular momentum operator, $\hbar\  \hat{\textbf{S}}_{z}=\frac{\hbar}{2} \sum\limits_{i=1}^{3} \ \sigma^{(i)}_{z}$. The above states are the eigenstates of the cyclic permutation operator denoted by $P_{+}$. \\
 
 As a first approximation, when the chemical shift anisotropy and the dipole-dipole interactions are negligible compared to the Zeeman interaction, the Hilbert space of these identical spins can be partitioned as a direct sum of two subspaces, $\mathcal{H}=\mathcal{H}_{A} \oplus \mathcal{H}_{E}$, and each subspace can be further decomposed into a product of the symmetry label $s$ and the magnetization label $m$, i.e., $\mathcal{H}_{A}= \mathbb{C}^1 \otimes \mathbb{C}^{4}$ and $ \mathcal{H}_{E}=\mathbb{C}^2 \otimes \mathbb{C}^{2}$. The first subsystem ($s$ label) is not affected by collective noise operators, because
 
 \begin{eqnarray}
 \hat{\textbf{S}}_{z} \ |s, m \rangle &=& m\  |s, m\rangle, \\ \nonumber
  \hat{\textbf{S}}_{x}\  |s, m \rangle &=& C_{+}\  |s, m+1 \rangle + C_{-}\  |s, m-1 \rangle, \\ \nonumber
    \hat{\textbf{S}}_{y} \ |s, m \rangle &=& C'_{+}\  |s, m+1 \rangle + C'_{-}\  |s, m-1 \rangle, \\ \nonumber
 \end{eqnarray}
 where $C_{\pm}$ and $C'_{\pm}$ are complex numbers. All components of the total spin angular momentum, $\hat{\textbf{S}}_{ \alpha}$ with $\alpha \in \{ x, y, z\}$, preserve the symmetry label although they may corrupt the magnetization label. One can take advantage of this symmetry property by polarizing the methyl group in terms of the symmetry order rather than the conventional Zeeman order. This way, the system is protected against collective spin noise and may exhibit long relaxation times. This has been experimentally demonstrated as \textit{Long Lived States} in methyl groups \cite{CL04} where an imbalance of population is created between the $A$ subspace and the $E_{\pm}$ subspaces. In the following, we provide a mathematical description for these noise protected states.\\
 
 We denote a state that is only populated in a particular symmetry subspace by $\rho_{s}$ with $s\in \{A, E_{+}, E_{-}\}$. These states are totally polarized in terms of the symmetry label, but they are totally mixed in terms of the magnetization label. Explicitly, we define
\begin{eqnarray}
\label{Eq_base_states}
\rho_{A} &:=& \frac{1}{4} \sum\limits_{m=-\frac{3}{2}}^{\frac{3}{2}} |A , m \rangle \langle A, m | = \frac{1}{4}\left(\begin{array}{c|c}
 \mathbb{1}_{4} & 0\\
 \hline
 0 & 0
\end{array} \right),\\\nonumber
\rho_{E_{+}} &:=& \frac{1}{2} \sum\limits_{m=-\frac{1}{2}}^{\frac{1}{2}} |E_{+} , m \rangle \langle E_{+}, m |=\frac{1}{2}\left(\begin{array}{c|c}
0 & 0\\
\hline
 0 & \begin{array}{c|c}
\mathbb{1}_{2} & 0\\
\hline
 0 & 0
\end{array}
\end{array} \right),\\ \nonumber
\rho_{E_{-}} &:=& \frac{1}{2} \sum\limits_{m=-\frac{1}{2}}^{\frac{1}{2}} |E_{-} , m \rangle \langle E_{-}, m |=\frac{1}{2}\left(\begin{array}{c|c}
0 & 0\\
\hline
 0 & \begin{array}{c|c}
0 & 0\\
\hline
 0 & \mathbb{1}_{2}
\end{array}
\end{array} \right).
\end{eqnarray}

With the above definitions, we introduce a $\gamma-$polarized long lived state as
 \begin{eqnarray}
 \label{Eq_LLS}
 Q_{LLS}&= &\frac{(1+\gamma)}{2}\rho_{A} + \frac{(1-\gamma)}{2} \left(\frac{\rho_{E_{+}}+ \rho_{E_{-}} } {2}\right)\\ \nonumber
 &=&\frac{ 1}{8}( \mathbb{1} +  \frac{ \gamma}{3}\ (\vec{\sigma}^{(1)}. \vec{\sigma}^{(2)} + \vec{\sigma}^{(2)}. \vec{\sigma}^{(3)}  + \vec{\sigma}^{(1)}. \vec{\sigma}^{(3)})).
 \end{eqnarray}
  The appearance of the scalar terms, $\vec{\sigma}. \vec{\sigma}$, assures that $Q_{LLS}$ is not affected by any component of the total spin magnetization, $\hat{\textbf{S}}_{\alpha}$, with $\alpha \in \{x,y,z\}$. One could extend this idea and define a more general type of protected states as

\begin{eqnarray}
 \label{Eq_Protected}
 Q_{\text{Protected}}=\frac{(1+\gamma)}{2}\rho_{A} + \frac{(1-\gamma)}{2}\left( \frac{1+ \beta}{2} \rho_{E_{+}} + \frac{1- \beta}{2} \rho_{E_{-}} \right) \\ \nonumber
 \end{eqnarray} \\
where the $A$ subspace is $\gamma$ polarized relative to the $E$ subspace and the $E_{+}$ subspace is $\beta$ polarized relative to the $E_{-}$ subspace.
In this report, we are not concerned about the initialization of the system in $Q_{LLS}$ or $Q_{\text{protected}}$. We assume that a long lived state is given and we study the dipolar relaxation mechanism that turns this noise protected state into an NMR observable.

\section{Symmetry of Dipolar Coupling}
\label{Sec_Dipolar}
So far, we have neglected that not all interaction terms in the Hamiltonian preserve the cyclic permutation symmetry. Ideally, if the first two terms of Eq.\ref{Eq_Spin_Ham} are the only terms in the Hamiltonian, spins are completely indistinguishable, and therefore, the long lived states can never be converted into an NMR observable, because, if they are protected against collective noise, they are protected against the measurement as well. Quantitatively, $Tr[Q_{LLS}\  \mathbf{S}_{\alpha}]  = 0$ for all $\alpha\in \{x,y,z\}$. However, the chemical shift anisotropy, the internal dipolar coupling among protons in a methyl group and the external dipolar coupling between the group of protons and an external spin (such as $^ {13}C$) are all possible interactions that do not preserve the spin symmetry. These physical interactions that distinguish spins can potentially transform the symmetry polarization into Zeeman polarization, leading to NMR observation of what were once protected states. The focus of our study is the contribution of dipole-dipole interaction in the spin read out of an initially prepared protected state. The CSA is normally small for protons in methyl groups in comparison with the dipolar coupling, and hence, is not considered in this study, although the symmetry arguments would be similar.

\subsection{Dipolar Hamiltonian}
\label{Sec_Dip_Ham}
Two magnetic dipole moments $\vec{d}_{1}$ and $\vec{d}_{2}$ that are at a distance $\vec{\textbf{r}}$ apart, interact through space via the dipolar Hamiltonian \cite{A62},  
 \begin{equation}
 \label{Eq_dipolar}
H_{\text{dip}}= - \frac{\mu_{0}}{4 \pi} \frac{1}{ r^{5}}\left( 3(\vec{d}_{1}.\vec{r})(\vec{d}_{2}.\vec{r}) - \vec{d}_{1}.\vec{d}_{2}\right).
\end{equation}
Here, $\vec{d}_{1}= \hbar \gamma_{S}\ \vec{S} $ and $\vec{d}_{2}= \hbar \gamma_{I}\ \vec{I}$ with $\vec{S}=\frac{1}{2}\vec{\sigma}$ and $\vec{I}=\frac{1}{2}\vec{\sigma}$ for the case of spin half particles where $\vec{\sigma}$ are Pauli operators. The use of different symbols $S$ and $I$ emphasizes that these operators act on different Hilbert spaces and may represent two different spin species. The dipolar Hamiltonian is commonly written in a product form of spatial functions and irreducible rank-2 tensors \cite{A62}, 
\begin{eqnarray}
H_{\text{dip}}&=& \sum\limits_{q=-2}^{2} H_{q}\\ \nonumber &=&
 c_{0}\ \sum\limits_{q=-2}^{2} \  e^{-i q\varphi}\ F_{q}( r, \theta) \  \hat{T}_{q},
\end{eqnarray}
where $c_{0}=-\frac{\hbar\mu_{0} \gamma_{I}\gamma_{S}}{4 \pi} $ is a constant, $F_{q}( r, \theta) $ a function of space parameters and $\hat{T}_{q}$ a normalized bilinear spin operator. Explicitly,
\begin{eqnarray}\nonumber
\hat{T}_{0}&=& \sqrt{\frac{2}{3}}\ \left( 3 S_{z}I_{z}- \vec{S}.\vec{I}\right), \hskip 0.5 cm  F_{0}= \sqrt{\frac{3}{2}}\frac{1}{r^{3}} \frac{(3 \cos^2\theta-1)}{2},\\ \nonumber
\hat{T}_{\pm}&=& \mp \left( S_{z}I_{\pm}+ S_{\pm}I_{z}\right), \hskip 0.6 cm  F_{\pm1}=  \mp \frac{3}{2} \frac{1}{r^{3}}\sin\theta \cos\theta, \\ \nonumber
\hat{T}_{\pm2}&=&  S_{\pm}I_{\pm}, \hskip 1.5 cm  F_{\pm 2}= \frac{3}{4} \frac{1}{ r^3}\sin^{2}\theta.
\end{eqnarray}
Given these definitions, we have
\begin{eqnarray}
\hat{T}_{q}^{\dagger} &=& (-1)^{q} \hat{T}_{-q}\\ \nonumber
 Tr[\hat{T}_{q}^{\dagger} \hat{T}_{q'}] &=& \delta_{q,q'},\\ \nonumber
  F_{-q} &=& (-1)^{q} F_{q}.
\end{eqnarray}

In case of the homonuclear interaction, since $[S_{z}+ I_{z}, \hat{T}_{q}] = q\ \hat{T}_{q}$, each $\hat{T}_{q}$ term changes the $z$ component of the total spin magnetization from $m$ to $m+ q$. So, we may refer to $q$ as the \textit{order} number. \\ 

In case of hetronuclear interactions, each $\hat{T}_{q}$ is further decomposed to $\hat{T}_{q} = \sum\limits_{p} \ \hat{T}_{(q,p)}$  based on its commutation with the Zeeman Hamiltonian, i.e., $[ \omega_{s} \ S_{z} + \omega_{I}\  I_{z} \ , \ \hat{T}_{(q,p)} ]= \omega_{(q,p)}\  \hat{T}_{(q,p)}$. The explicit form of these bilinear spin operators and their corresponding frequencies are listed in Table. \ref{Tab_dipolar_component}. One can write them in a closed form by noting $\hat{T}_{(q,p)}\propto  S_{p} \otimes I_{q-p}$ and $\omega_{(q,p)}= p\ \omega_{s} + (q-p)\ \omega_{I}$, where $I_{0}$ or $S_{0}$ represents $\frac{\sigma_{z}}{2}$, and $I_{\pm}$ or $S_{\pm}$ represents $\frac{\sigma_{x} \pm i \sigma_{y}}{2}$. 
\begin{table}[h]
\begin{center}
\begin{tabular}{c|ccc| ccc}
$q$& & $\hat{T}_{(q,p)}$ & & & $\omega_{(q,p)}$& \\
\hline
  &  $p=1$ & $p=0$ &  $p=-1$ & $p=1$ & $p=0$ &  $p=-1$  \\
\hline \hline
0 &  -$\frac{1}{\sqrt{6}}S_{+}I_{-}$  & $\sqrt{\frac{8}{3}} S_{z} I_{z}$& $-\frac{1}{\sqrt{6}}S_{-}I_{+}$ &  $\omega_{s}- \omega_{I}$ & 0 & $-\omega_{s}+ \omega_{I}$ \\ 
+1 &   $S_{+} I_{z}$ & $S_{z}I_{+}$ & --- & $\omega_{s}$ & $\omega_{I}$& ---  \\
+2 &  $S_{+} I_{+}$ & --- & --- &$\omega_{s}+ \omega_{I}$ &--- & ---
\end{tabular}
 \caption{Components of the Dipolar Hamiltonian.}
 \label{Tab_dipolar_component}
\end{center}
\end{table} 

Given the above definitions, the homonuclear coupling between protons of a methyl group is

\begin{eqnarray}
\label{Eq_homo_dipolar}
H^{SS}_{\text{dip}}&=&  H_{\text{dip}}^{(1,2)}+ \ H_{\text{dip}}^{(2,3)}+  H_{\text{dip}}^{(3,1)}\\\nonumber
&=&   c_{1}\sum\limits_{i<j} \sum\limits_{q=-2}^{2} \  e^{-i q\varphi_{ij}}\ F_{q}( r_{ij}, \theta_{ij}) \  \hat{T}^{(i,j)}_{q}\\
&=&\sum\limits_{q=-2}^{2} H^{SS}_{q},
\end{eqnarray}
where $H_{\text{dip}}^{(i,j)}$ represents the dipole-dipole coupling between the $i^{\text{th}}$ proton and the $j^{\text{th}}$ proton and is given in Eq.\ref{Eq_dipolar} with $S= S^{i}$ and $I=S^{j}$.  In addition, $\vec{\textbf{r}}_{ij}= (r_{ij}, \theta_{ij}, \varphi_{ij})$ is the relative distance between the two protons and $c_{1}=-\frac{\hbar\mu_{0} \gamma_{I}\gamma_{I}}{4 \pi} $ is a constant.\\

Similarly, the hetronuclear coupling between the collective spins, $\mathbf{S}$, and the test spin, $I$, consists of three terms,
\begin{eqnarray}
\label{Eq_hetro_dipolar}
H^{\mathbf{S}I}_{\text{dip}}&=&  H_{\text{dip}}^{(1)}+ \ H_{\text{dip}}^{(2)}+  H_{\text{dip}}^{(3)}\\\nonumber
&=&   c_{0}\sum\limits_{j=1}^{3} \sum\limits_{q,p} \  e^{-i q\varphi_{j}}\ F_{q}( r_{j}, \theta_{j}) \  \hat{T}^{j}_{(q,p)}\\
&=&\sum\limits_{q}\  H^{\mathbf{S}I}_{q},
\end{eqnarray}
where $H_{\text{dip}}^{(j)}$ represents the dipole-dipole coupling between the $j^{\text{th}}$ proton, $S^{j}$ and the test spin, $I$, and $\vec{\textbf{r}}_{j}= (r_{j}, \theta_{j}, \varphi_{j})$ is the relative distance between them in the Zeeman frame.\\

The hetronuclear dipolar Hamiltonian does not commute with the spin cyclic permutation operator, i.e., $[P_{+}\ ,\ \sum\limits_{q}\ H^{\mathbf{S}I}_{q} ]\neq 0$. Similarly, the homonuclear dipolar Hamiltonian does not commute either. This non-commutativity implies that the dipolar Hamiltonian and the cyclic permutation operator, $\hat{\textit{P}}_{+}$, do not have a common basis, and hence, the dipolar interaction can induce transitions between the eigenstates with different symmetry labels. Consequently, a dipolar relaxation might lead to a process that takes polarization from symmetry order to Zeeman order, which is measurable.
\subsection{Symmetrized Operators}
\label{Sec_Symm_Op}
In the following, we introduce the symmetrized bilinear spin operator and re-write the dipolar Hamiltonian in terms of them. These symmetrized operators provide insight about the key components of the dipolar coupling that leads to observing a protected state. We start with the external dipolar interaction and then elaborate on the internal dipolar coupling and discuss their differences and similarities.\\

For hetronuclear coupling, we substitute $B^{j}_{q}  = c_{0} \ e^{-i q \varphi_{j}} \ F_{q}(r_{j}, \theta_{j}) $ in Eq.\ref{Eq_hetro_dipolar} and re-write the Hamiltonian as \cite{M76, BarB88}, 
 \begin{eqnarray}
H^{\mathbf{S}I}_{\text{dip}}(t) =\sum\limits_{q,p}&& \sum\limits_{j=1}^{3}\ B^{j}_{q}(t) \ \hat{T}_{(q,p)}^{j} \\ \nonumber
= \sum\limits_{q,p}\{  +  \frac{1}{3} &&\ \left( B^{1}_{q} +  \ B^{2}_{q} + \ B^{3}_{q}\right) \\ \nonumber &\times & \left( \hat{T}^{1}_{(q,p)}+  \hat{T}^{2}_{(q,p)} + \ \hat{T}^{3}_{(q,p)}\right)\\ \nonumber
+ \frac{1}{3} &&\ \left( B^{1}_{q} +  \epsilon^{*} \ B^{2}_{q} + \epsilon \ B^{3}_{q}\right) \\ \nonumber &\times & \left( \hat{T}^{1}_{(q,p)} + \epsilon\   \hat{T}^{2}_{(q,p)} + \epsilon^{*}\ \hat{T}^{3}_{(q,p)}\right)\\ \nonumber
+ \frac{1}{3}&&\ \left( B^{1}_{q} +  \epsilon \ B^{2}_{q} + \epsilon^{*} \ B^{3}_{q}\right) \\ \nonumber &\times & \left( \hat{T}^{1}_{(q,p)} + \epsilon^{*}\   \hat{T}^{2}_{(q,p)} + \epsilon\ \hat{T}^{3}_{(q,p)}\right) \}\\ \nonumber
= \sum\limits_{q,p}&& \sum\limits_{\lambda=0, \pm1} \ B_{q}^{\lambda}(t) \ \hat{T}_{(q,p)}^{\lambda},
 \end{eqnarray}
where $\epsilon= e^{\frac{i 2 \pi}{3} }$. We emphasize that in the last line, the sum over different spin indices is replaced by a sum over the symmetry label $\lambda\in\{ 0 , \pm 1\}$, which corresponds to $\{A, E_{\pm}\}$ symmetries. The symmetrized complex functions $ B^{\lambda}_{q}$ and the symmetrized bilinear spin operator  $\hat{T}^{\lambda}_{(q,p)}$ are given by
 \begin{eqnarray}
   B^{\lambda}_{q}&:=& \frac{1}{\sqrt{3}}\ \left( B^{1}_{q} + \epsilon^{\lambda*} \ B^{2}_{q} +  \epsilon^{\lambda} \ B^{3}_{q}\right), \\ \nonumber
  \hat{T}^{\lambda}_{(q,p)}&:=& \frac{1}{\sqrt{3}}\ \left( \hat{T}^{1}_{(q,p)} +  \epsilon^{\lambda}\ \hat{T}^{2}_{(q,p)} +  \epsilon^{\lambda*} \ \hat{T}^{3}_{(q,p)}\right). 
 \end{eqnarray}
  In closed form, $\hat{T}^{\lambda}_{(q,p)} \propto \textbf{S}_{p}^{\lambda} \otimes I_{q-p}$, where the bold notation reminds us that $\textbf{S}_{p}^{\lambda}$ acts on all three spins.\\
  
   To clarify the effect of $\textbf{S}_{p}^{\lambda}$ on the collective spin system, consider a simpler example of two spins, where $\epsilon= e^{\frac{i 2\pi}{2}}= -1$ and $\lambda \in\{0 ,1\}$ (or $\{A,E\}$). Thus, $\textbf{S}_{p}^{A/E} \propto \left( S^{1}_{p} \pm S^{2}_{p} \right)$ is either symmetric or anti-symmetric. By looking at the non-zero matrix elements of $\textbf{S}_{p}^{\lambda}$ in the triplet-singlet basis, we conclude that the symmetric tensors $\textbf{S}^{A}_{p}$ causes transitions only \textit{within} each symmetry subspace and the anti-symmetric tensors $\textbf{S}^{E}_{p}$ causes transitions \textit{between} two subspaces of the triplet states ($A$) and  the singlet state ($E$).  Indeed, for two spins we obtain
  \begin{equation}
 \textbf{S}^{\lambda}_{p}= \frac{1}{\sqrt{2}}\left( S^{1}_{p} + (-1)^{\lambda} S^{2}_{p} \right)\ | j, m\rangle \longrightarrow | j + \lambda , m + p\rangle
  \end{equation}
  where $j=1$ is the triplet subspace and $j=0$ is the singlet subspace. 
 \begin{figure}[h]
  \centering
\includegraphics[width=0.8\linewidth]{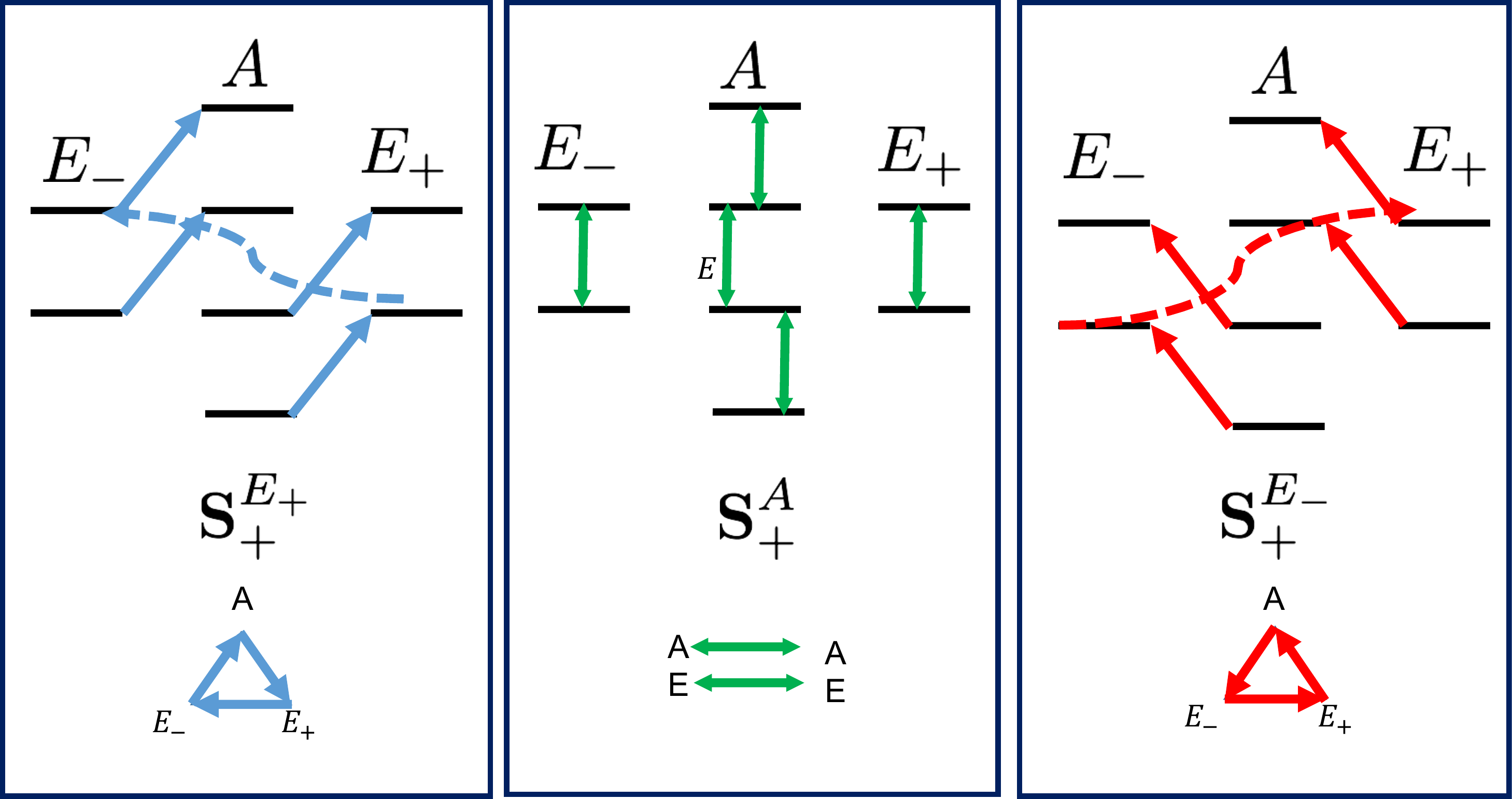}
  \caption[Allowed Transitions due to $\mathbf{S}_{+}^{\lambda}$]{Allowed Transitions due to $\mathbf{S}_{+}^{\lambda}$: The non-zero transitions between the eigenstates of $P_{+}$ are shown that are due to the non-zero matrix elements of the symmetrized collective spin operators $\mathbf{S}_{+}^{\lambda}$. The lower index $+$ acts on the magnetization label where it takes $m$ to $m+1$ and the upper index $\lambda$ acts on the symmetry label in a cyclic manner. The blue/red arrows indicate the non-zero transitions between different symmetry spaces in the right/left cyclic order and the green arrows refer to non-zero transition within each symmetry space.}
  \label{fig_Cyclic_Transitions}
\end{figure}

  Extending this to three spins is straightforward. Similarly, the $\textbf{S}^{A}_{p}$ of three spins is a totally symmetric operator and it has non-zero matrix elements only within each symmetry subspaces $A$, $E_{+}$ or $E_{-}$. Consequently, it is block-diagonal in the eigensbasis of the cyclic permutation operator.  In a similar manner to the two-spin case, $\textbf{S}^{E_{\pm}}_{p}$ links two different symmetry subspaces. But, in case of three spins, $\textbf{S}^{E_{\pm}}_{p}$ is neither symmetric nor anti-symmetric and one needs to be careful about the direction of transitions. We expand the $\textbf{S}^{E_{\pm}}_{p}$ in the $|s, m\rangle$ basis and conclude that $\textbf{S}^{E_{+}}_{p}$ transforms the symmetry of the eigenstates in \textit{right cycle} as $(A\rightarrow E_{+}\rightarrow E_{-}\rightarrow A)$ and the $\textbf{S}^{E_{-}}_{p}$ transforms the symmetry of the eigenstates in \textit{left cycle} as $(A\leftarrow E_{+}\leftarrow E_{-}\leftarrow A)$. In summary, the upper index of $\textbf{S}^{\lambda}_{p}$ determines whether the transformation is \textit{within} each symmetry subspace $(\lambda=A)$ or \textit{between} them in a right/left cyclic direction ($\lambda= E_{\pm})$. And, the lower index determines the change in the the magnetization. Explicitly, when $p=0$, $\textbf{S}^{\lambda}_{0}$ takes $m\rightarrow m$ and when $p=\pm1$ the $\textbf{S}^{\lambda}_{\pm }$ takes $m\rightarrow m \pm 1$ for all values of $\lambda$. In other words, 
   \begin{equation}
 \textbf{S}^{\lambda}_{p}\ | s, m\rangle \longrightarrow | s + \lambda , m + p\rangle
  \end{equation}
  where the sum in $s + \lambda$ is mod 3.  The final remark is that the reverse transition process occurs via $(\textbf{S}^{\lambda}_{p})^{\dagger}$, where
  \begin{equation}
  (\mathbf{S}_{p}^{\lambda})^{\dagger} = \mathbf{S}_{-p}^{-\lambda}.
  \end{equation}
  For example, the transition from $|E_{+}, \frac{1}{2}\rangle$ to $|A, \frac{3}{2}\rangle$ occurs through $\mathbf{S}_{+}^{E_{+}}$ but the reverse process, from $|A, \frac{3}{2}\rangle$ to $|E_{+}, \frac{1}{2}\rangle$ occurs through $\mathbf{S}_{-}^{E_{-}}$. The allowed transitions due to $\mathbf{S}_{+}^{\lambda}$ are demonstrated in Fig.\ref{fig_Cyclic_Transitions}.\\
  
The homonuclear dipolar Hamiltonian can also be written as a sum over symmetrized operators. We substitute $G^{ij}_{q}  = c_{1} \ e^{-i q \varphi_{ij}} \ F_{q}(r_{ij}, \theta_{ij}) $ in Eq.\ref{Eq_homo_dipolar} and re-write the Hamiltonian as, 
 \begin{eqnarray}
H^{SS}_{\text{dip}}(t) =\sum\limits_{q,p}&& \sum\limits_{i<j}^{3}\ G^{ij}_{q}(t) \ \hat{T}_{q}^{(i,j)} \\ \nonumber
= \sum\limits_{q,p}&& \sum\limits_{\lambda=0, \pm1} \ G_{q}^{\lambda}(t) \ \hat{\mathbb{T}}_{(q,p)}^{\lambda},
 \end{eqnarray}
  where  
   \begin{eqnarray}
   G^{\lambda}_{q}&:=& \frac{1}{\sqrt{3}}\ \left( G^{12}_{q} + \epsilon^{\lambda*} \ G^{23}_{q} +  \epsilon^{\lambda} \ G^{31}_{q}\right), \\ \nonumber
  \hat{\mathbb{T}}^{\lambda}_{q}&:=& \frac{1}{\sqrt{3}}\ \left( \hat{T}^{(1,2)}_{q} +  \epsilon^{\lambda}\ \hat{T}^{(2,3)}_{q} +  \epsilon^{\lambda*} \ \hat{T}^{(3,4)}_{q}\right). 
 \end{eqnarray}
 
\begin{figure}[h]
  \centering
\includegraphics[width=0.8\linewidth]{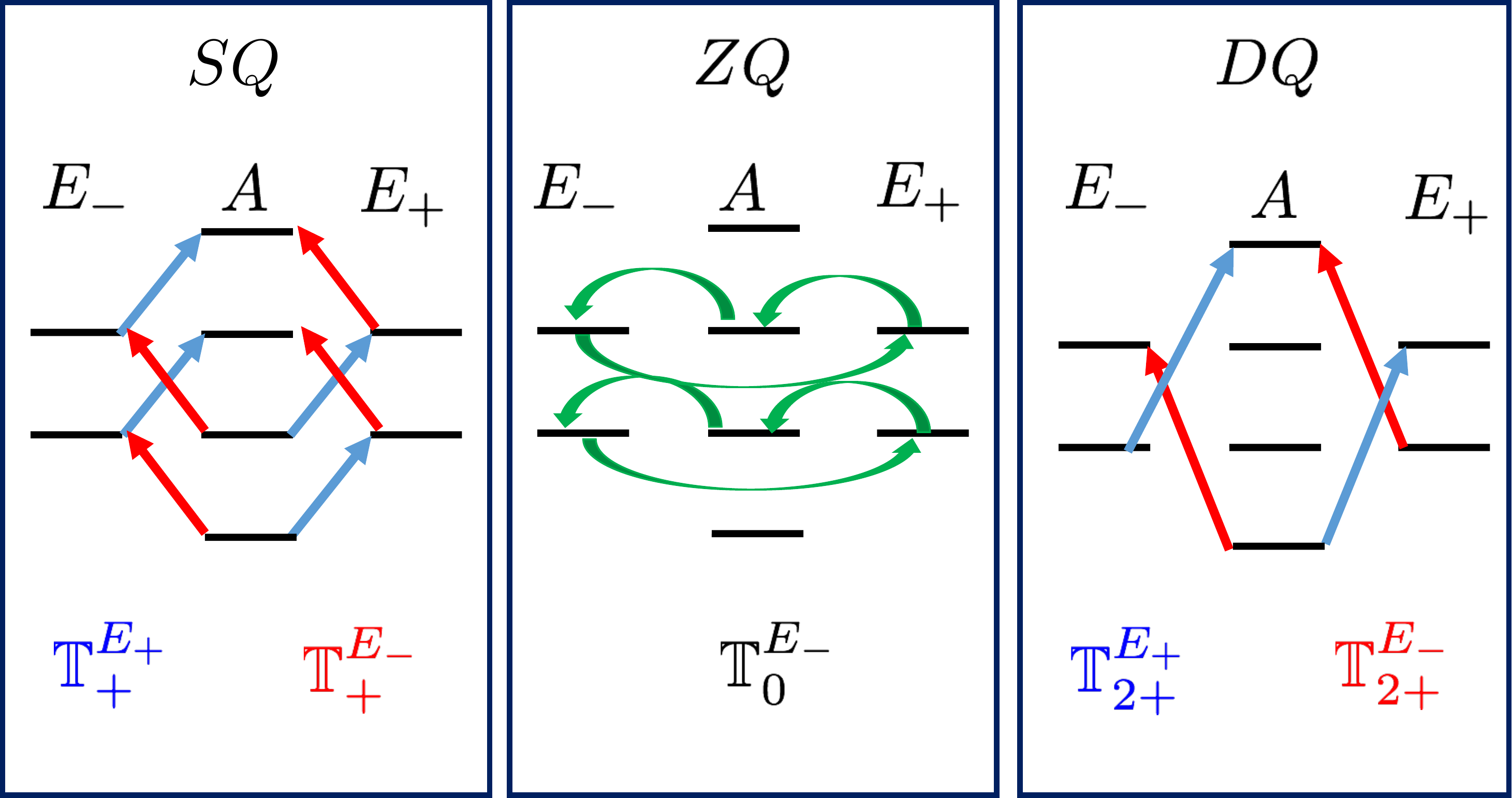}
  \caption[Allowed Transitions due to $\mathbb{T}_{q}^{\lambda}$]{Allowed Transitions due to  $\mathbb{T}_{q}^{\lambda}$: The non-zero transitions between the eigenstates of $P_{+}$ are shown that are due to the non-zero matrix elements of the symmetrized bilinear spin operators  $\mathbb{T}_{q}^{\lambda}$. From left to right, the figure shows allowed transitions due to the SQ ($q=1$), the ZQ ($q=0$) and the DQ ($q=2$) spin operators. The blue/red arrows indicate the non-zero transitions between different symmetry spaces in the right/left cyclic. The green arrows indicate those transitions that do not effectively contribute in the NMR signal.}
\label{fig_Homo_Transitions} 
\end{figure}

 We observe that the symmetrized spin operators for internal dipolar coupling, $\mathbb{T}_{q}^{\lambda}$, exhibit similar symmetry properties to that of external dipolar coupling, $\mathbf{S}_{p}^{\lambda}$. Similarly, the fully symmetrized operators, $\lambda =0$ or $A$, have non-zero matrix elements only \textit{within} each symmetry subspace and not between them, and hence, they are not of interest. Thus, the important terms are $\lambda = E_{\pm}$ and Fig.\ref{fig_Homo_Transitions} shows all the non-zero transitions induced by $\mathbb{T}_{q}^{\lambda}$ for $q=0,1,2$. Note that in the figure the $S_{z}^{i} S^{j}_{z}$ terms are not considered because they contribute in shifting the energy levels not in the transitions.

The above transitions can be categorized in three groups: the zero quantum transitions (ZQ) also known as \textit{ flip-flop} terms when $q=0$ and $m \rightarrow m$, the single quantum transitions (SQ) or \textit{single spin flip} terms when $q=1$ and $m \rightarrow m \pm1$, and  the double quantum transitions (DQ) or the \textit{flop-flop} terms when $q=2$ and $m \rightarrow m \pm2$.  Similar to the $\mathbf{S}_{\pm}^{\lambda}$ operators in case of hetronuclear Hamiltonian, the internal flip-flop interaction among protons also leads to transition between states with different symmetry in a cyclic order (left /right for $\lambda =E_{\mp}$).  More precisely, 
\begin{eqnarray}
\hat{ \mathbb{T}}_{0}^{\lambda} \ | s\, \ m\rangle \rightarrow |s+ \lambda\ ,\ m\rangle \hskip 1 cm  (A\rightleftarrows E_{+}\rightleftarrows  E_{-}\rightleftarrows  A)
\end{eqnarray}

These transitions are \textit{passive} in the sense that even though they lead to exchange of population between different symmetry subspaces but they will not effectively contribute in transferring polarization from the symmetry order to Zeeman order as it will be explained in the next section.\\

The $SQ$ and the $DQ$ transitions, the $\hat{ \mathbb{T}}_{+1}^{\lambda}$ and the $\hat{ \mathbb{T}}_{+2}^{\lambda}$ operators, are slightly different. They do induce transitions between the $A$ and the $E_{\pm}$ subspaces but they do not allow transition between the $E_{+}$ and the $E_{-}$ subspaces as opposed to the external dipolar coupling. For $q= 1 ,2$ we have

\begin{eqnarray}\nonumber
\hat{\mathbb{T}}_{q}^{E_{+}}:  | A\ ,\  m\rangle &\rightleftarrows& |E_{\pm} \ , \ m +q \rangle \hskip 1 cm A\rightleftarrows E_{\pm}\\ \nonumber
\hat{\mathbb{T}}_{q}^{E_{+}}:  | A\ ,\  m\rangle &\rightleftarrows& |E_{\pm} \ , \ m +q \rangle \hskip 1 cm A\rightleftarrows E_{\pm}
\end{eqnarray} 

Now that we visualized the effect of these symmetrized irreducible tensors on the symmetrized eigenbasis, we proceed to solve the master equation in the next section.
\section{Master Equation}
\label{Sec_MasterEq}
At high temperature and in liquid phase, the space coordinates $\vec{\textbf{r}}_{j}(t)= (r_j, \theta_{i}, \varphi_{i})$ randomly fluctuate in time, so does the dipolar Hamiltonian. In the absence of CSA, the total Hamiltonian consists of a Zeeman term $H_{0} = \omega_{h} \ \hat{\mathbf{S}}_{z} + \omega_{I} \ \hat{I}_{z}$, a scalar coupling $H_{\text{scalar}}= 2\pi\ (J_{0} \ \sum\limits_{ij} \vec{\sigma}^{i}. \vec{\sigma}^{j} + J_{1} \ \vec{\mathbf{S}}.\vec{I})$, and a fluctuating dipolar term $ H^{\mathbf{S}I}_{\text{dip}}(t)$ and $ H^{SS}_{\text{dip}}(t)$. The scalar coupling is totally symmetric and does not contribute to breaking the symmetry of the protected states, so, it is neglected in the following discussion, unless otherwise stated. In the following discussion, we analyze one of the two types of dipolar relaxation process (homo vs hetro) in the absence of the other. Of course, in the experiment, these two processes are not independent of each other and the evolution of one affects the other, because their corresponding Hamiltonians do not commute. But, for a short time period, $\delta t$, one may consider them as two independent processes. We first analyze the pure contribution of hetronuclear coupling in the NMR observation of initially prepared protected states and then we analyze the contribution of the homonuclear coupling and discuss their differences and similarities.\\
\subsection{Relaxation Induced by Hetronuclear Coupling}

At high field, we treat $ H^{\mathbf{S}I}_{\text{dip}}(t)$ as a perturbation term and apply Redfield's semi-classical theory \cite{A62, BP02} to study the collective spin dynamics.  Then, we solve the master equation analytically and predict the NMR signal.\\

 The Lindbladian form of the semi-classical master equation gives us \cite{BP02}
\begin{eqnarray}
\label{Eq_master_dipolar}\nonumber
\frac{\partial \tilde{\rho}}{\partial t} = \sum\limits_{\lambda} \sum\limits_{q,p} J^{\lambda}_{q}(\omega_{(q,p)}) \hat{T}_{(q,p)}^{\lambda}\ \tilde{\rho}\  \hat{T}_{(q,p)}^{\lambda\dagger} - \frac{1}{2}\{ \hat{T}_{(q,p)}^{\lambda\dagger}\hat{T}_{(q,p)}^{\lambda},\ \tilde{\rho}\} , 
\end{eqnarray}
where $\tilde{\rho}= e^{iH_{0} t} \rho e^{-iH_{0}t}$ is the density matrix in the rotating frame of the Zeeman interaction. The coefficient $J^{\lambda}_{q}(\omega)$ is the real part of the symmetrized spectral density of noise and is defined by
\begin{eqnarray}
J^{\lambda}_{q}(\omega)&:=& \int\limits_{-\infty}^{\infty} R_{q}^{\lambda}(\tau) \ e^{-i \omega\tau} \ d\tau , \\ \nonumber
R_{q}^{\lambda}(\tau) &:= &\overline{ B^{\lambda}_{q}(t) (B^{\lambda}_{q})^{*}(t+\tau)}.
\end{eqnarray}
Here, $R_{q}^{\lambda}(\tau)$ denotes the symmetrized auto correlation function and the overbar notation refers to averaging over the random variables. The imaginary part of the spectral density of noise leads into the dynamical shift and can be absorbed in the coherence evolution part \cite{KM06}.\\
   
\textit{For the purpose of the following discussion, the explicit form of the $J(\omega)$ is not required, because we are not interested in the exact dynamics of the system. Rather we would like to know which components of the hetronuclear dipolar coupling converts the symmetry polarization of the $Q_{LLS}$  into the Zeeman polarization which is measurable.} \\ 

We define the \textit{Lindbladian map} with $ \hat{D}[L][.]:= \hat{L}\ .\  \hat{L}^{\dagger} - \frac{1}{2}\{ \hat{L}^{\dagger}\hat{L},\ .\}$ in which $L$ is the Lindblad operator and re-write the master equation as

\begin{equation}
\frac{\partial \tilde{\rho}}{\partial t} = \sum\limits_{\lambda} \sum\limits_{q,p} J^{\lambda}_{q}(\omega_{(q,p)})\ \eta_{q,p} \ \hat{D}[ \textbf{S}_{p}^{\lambda} \otimes I_{q-p}][\ \tilde{\rho}]. 
\end{equation}
 where $\eta_{0.0}=8/3$, $\eta_{0,\pm} =-1/6$ and $\eta_{q,p}=1$ for all $q\neq 0$ cases. 
 Considering classical noise $ J^{\lambda}_{q}(\omega) = J^{-\lambda}_{-q}(-\omega)$ and neglecting the $q=p=0$ term which just shifts the energy, one can break the above master equation into three parts: the zero quantum transitions (ZQ), the double quantum transitions (DQ) and the single quantum transitions (SQ), 
 \begin{eqnarray}
 \label{Eq_master_dipolar2}
 \frac{\partial \tilde{\rho}}{\partial t} &=& \sum\limits_{\lambda} \  ZQ^{\lambda}[\tilde{\rho}] +  DQ^{\lambda}[\tilde{\rho}] +  SQ^{\lambda}[\tilde{\rho}] \\ \nonumber
 ZQ^{\lambda}[.]&:=& \frac{1}{6}\ J^{\lambda}_{0}(\omega_{s}- \omega_{I})\  \left( \hat{D}[ \textbf{S}_{+}^{\lambda} \otimes I_{-}][.] + \hat{D}[ \textbf{S}_{-}^{-\lambda} \otimes I_{+}][.] \right) \\ \nonumber
  DQ^{\lambda}[.]&:=&  J^{\lambda}_{2}(\omega_{s} +\omega_{I})\  \left( \hat{D}[ \textbf{S}_{+}^{\lambda} \otimes I_{+}][.] + \hat{D}[ \textbf{S}_{-}^{-\lambda} \otimes I_{-}][.] \right) \\ \nonumber
   SQ^{\lambda}[.]&:=& J^{\lambda}_{1}( \omega_{s}) \ \left( \hat{D}[ \textbf{S}_{+}^{\lambda} \otimes I_{z}][.] + \hat{D}[ \textbf{S}_{-}^{-\lambda} \otimes I_{z}][.]  \right) \\ \nonumber &+& J^{\lambda}_{1}( \omega_{I}) \ \left( \hat{D}[ \textbf{S}_{z}^{\lambda} \otimes I_{+}][.] + \hat{D}[ \textbf{S}_{z}^{-\lambda} \otimes I_{-}][.]  \right) .
 \end{eqnarray}
 Here, we replace $ (\mathbf{S}_{p}^{\lambda})^{\dagger} = \mathbf{S}_{-p}^{-\lambda}$. The ZQ and the DQ terms exchange energy between the collective spin and the test spin, and the SQ terms change either the collective spin states or the test spin states. As was mentioned before, the totally symmetric Lindblad operators $\textbf{S}_{p}^{A} \otimes I_{q-p}$ do not cause transition between two different symmetry subspaces of the collective spin. Therefore, if it happens that the spectral density of noise is very well approximated with just the totally symmetric component, i.e., $J^{\lambda}_{q}(\omega) \approx J^{A}_{q}(\omega) $, one can conclude that the system is very robust against noise and exhibits very long relaxation time. This is in agreement with the result in \cite{M13}. For the sake of simplicity in the following discussion, we ignore all of the totally symmetric Lindblad operators, since they do not play a critical role in observing the protected state. We also neglect the $SQ^{\lambda}[.]$ terms, because we are interested in those transitions that the collective spin exchanges the energy with the test spin. The only important terms in the dissipator are $DQ^{E_{\pm}}$ and $ZQ^{E_{\pm}}$.\\

To obtain the allowed transitions due to ZQ and DQ terms, we need to calculate the effect of $\hat{D}[ \textbf{S}_{p}^{\lambda} \otimes I_{\pm}] [.]$ on the eigenbasis $\{ |s, m\rangle \otimes |\uparrow \rangle \text{or} |\downarrow\rangle\}$. The non-zero components are,
\begin{widetext}
\begin{eqnarray}
\label{Eq_Master_To_Rate}
\hat{D}[ \textbf{S}_{p}^{\lambda} \otimes I_{-}][|s,m\rangle \langle s,m| \otimes | \uparrow \rangle \langle \uparrow| ] =&&  \textbf{S}_{p}^{\lambda}\ |s,m\rangle \langle s,m| \ \textbf{S}_{p}^{\lambda})^{\dagger} \otimes I_{-}\ | \uparrow \rangle \langle \uparrow|\  I_{+}\\ \nonumber
&&  - \frac{1}{2}\left(\textbf{S}_{-p}^{-\lambda}\textbf{S}_{p}^{\lambda} \ |s,m\rangle \langle s,m|\  \otimes  I_{+}I_{-} \ | \uparrow \rangle \langle \uparrow| \right)\\ \nonumber
&& - \frac{1}{2}\left( |s,m\rangle \langle s,m| \ \textbf{S}_{-p}^{-\lambda}\textbf{S}_{p}^{\lambda} \otimes   | \uparrow \rangle \langle \uparrow|\  I_{+}I_{-} \right)\\ \nonumber
&\propto & |s+\lambda, m+p \rangle \langle s+\lambda, m+p| \otimes | \downarrow \rangle \langle \downarrow |- |s, m\rangle \langle s, m| \otimes  | \uparrow \rangle \langle \uparrow|.
\end{eqnarray} 

 Similarly, 
 $$\hat{D}[ \textbf{S}_{p}^{\lambda} \otimes I_{+}][|s,m\rangle \langle s,m| \otimes | \downarrow \rangle \langle \downarrow| ] \propto  |s+\lambda, m+p \rangle \langle s+\lambda, m+p| \otimes | \uparrow \rangle \langle \uparrow |- |s, m\rangle \langle s, m| \otimes  | \downarrow \rangle \langle \downarrow|.$$
 \end{widetext}
The proportionality constant is $1$ for transitions from or to the $m=\pm3/2$ and is $\frac{1}{3}$ for all other levels. \\

\begin{figure}[h]
  \centering
\includegraphics[width=0.9\linewidth]{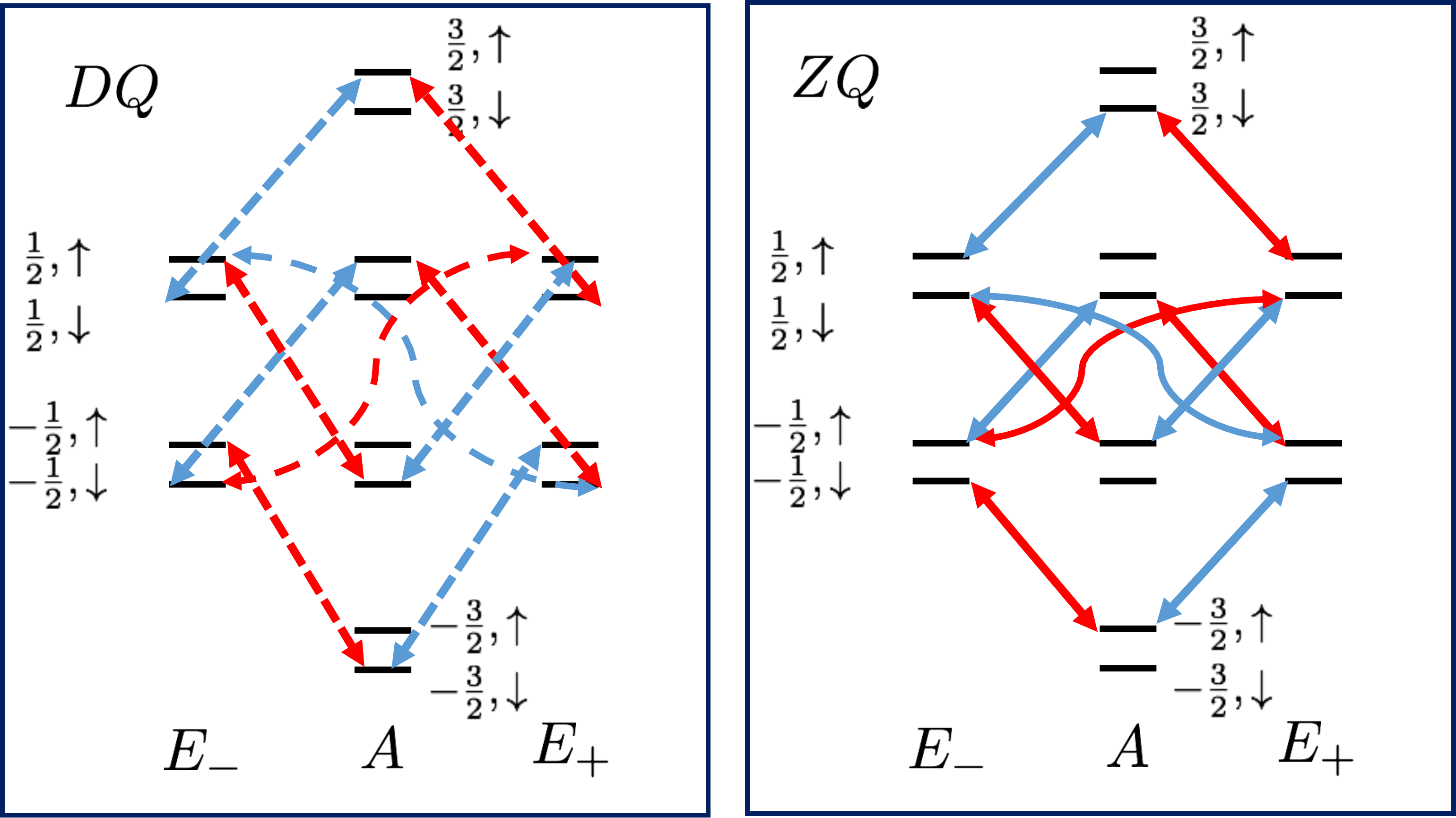}
  \caption[Double Quantum and Zero Quantum Allowed Transitions]{ \label{fig_non_zero_transitions}Left segment: The selection rule due to $DQ^{E_{\pm}}$ transitions. Right segment: The selection rule due to $ZQ^{E_{\pm}}$ transitions. Red vs blue refers to non-zero transitions due to $\mathbf{S}^{E_{-}}_{p}$ versus  $\mathbf{S}^{E_{+}}_{p}$.}
\end{figure}

Based on the above relations, the allowed transition due to $ZQ^{E_{\pm}}$ and $DQ^{E_{\pm}}$ terms are indicated in Fig.\ref{fig_non_zero_transitions} in which $\lambda = E_{\pm}$ transitions are color coded with red and blue respectively. 

 The calculation in Eq.\ref{Eq_Master_To_Rate} convinces us that if the initial state is a probabilistic mixture of different energy levels, the above master equation reduces to the classical rate equations,
\begin{equation}
\label{Eq_rate_eq}
\frac{d}{dt}p_{x}(t) = \sum\limits_{y\neq x} W_{xy}\  \left(p_{y}(t)- p_{x}(t)\right),
\end{equation}
where $p_{x}(t)$ is the population of the $x^{\text{th}}$ energy level at time $t$ with $x \in \{ |s,m\rangle \otimes ( | \uparrow\rangle \text{or} | \downarrow\rangle ) \}$, and $W_{xy}$ is the transition rate between two energy levels $x$ and $y$. To solve the above differential equation, the evolution time is discretized into $N$ steps where $t_{N}= N\delta t$ and $\delta t$ is small compared to the energy scales of the system. Given the population distribution at time $t_{n}$, the change in the population of $x^{\text{th}}$ level at time $t_{n+1}$ is approximated by $\Delta p_{x}(n+1)\approx \delta t \times \sum\limits_{xy} \ W_{xy}\ \left( p_{y}(n) - p_{x}(n)\right)$. Therefore, if the initial condition is known, one should be able to calculate the NMR signal by solving the rate equations recursively. Note that at any instance of time, $t= t_{0}+ \delta t$ the transition occurs only between those energy levels that first, are allowed due to $ZQ^{E_{\pm}}$ and $DQ^{E_{\pm}}$terms, and second, there is an imbalance of population at $t=t_{0}$.\\

The initial state of interest is
\begin{eqnarray}
\label{Eq_Initial_state}
\rho_{0}&=& Q_{\text{Protected}} \otimes \left( \frac{ \mathbb{1}}{2} + \alpha\  I_{z} \right), \\ \nonumber
 &=& \left(\begin{array}{c|c|c}
\frac{1+\gamma}{2} \frac{\mathbb{1}}{4} &0 &0\\
\hline
0& \frac{1- \gamma}{2} \frac{1+\beta}{2}\frac{\mathbb{1}}{2} &0 \\
\hline
0& 0& \frac{1- \gamma}{2} \frac{1-\beta}{2}\frac{\mathbb{1}}{2} \\ 
\end{array}\right) \otimes \left( \begin{array}{cc}
\frac{1+\alpha}{2} &0\\
0& \frac{1-\alpha}{2} \\
\end{array}\right)
\end{eqnarray}

\noindent 
where $\alpha$ is the polarization of the the test spin and $Q_{\text{Protected}}$ is replaced from Eq.\ref{Eq_Protected}. Since $\rho_{0}$ is diagonal, we alternatively represent it by a \textit{population vector}
\begin{eqnarray}
\vec{p}(0) = \begin{array}{c} \uparrow\\  \downarrow \\ \end{array}
\left(\begin{array}{c} \frac{1+\alpha}{2} \vec{q}_{0} \\ \hline \frac{1-\alpha}{2} \vec{q}_{0}\\ \end{array}\right)
\end{eqnarray}
where
\begin{eqnarray}
\vec{q}_{0}= 
\left(\begin{array}{c}
\frac{(1+\gamma)}{2} \frac{1}{4}
 \left(\begin{array}{c}
  1\\ 
  1 \\
  1\\
  1\\
  \end{array}\right)\\
 \hline
 \frac{(1-\gamma)}{2} \frac{(1+\beta)}{2} \frac{1}{2} \left(\begin{array}{c}
  1\\ 
  1 \\
  \end{array}\right) \\
 \hline
 \frac{(1-\gamma)}{2} \frac{(1-\beta)}{2} \frac{1}{2} \left(\begin{array}{c}
  1\\ 
  1 \\
  \end{array}\right) \\
\end{array}\right) \small \begin{array}{ cc}
%
%
 A,\frac{3}{2}& \\
A,\frac{1}{2}& \\
A, -\frac{1}{2}&  \\
A, -\frac{3}{2} & \\

E_{+}, \frac{1}{2}& \\
E_{+}, -\frac{1}{2}& \\

E_{-}, \frac{1}{2} &\\
E_{-}, -\frac{1}{2}&\\
\end{array}.
\normalsize
 \end{eqnarray}
The $\vec{q}_{0}$ represents the vector of population of three identical spins at $t=0$ in which all energy levels with the same symmetry label are equally populated, but, there is an imbalance of population between energy levels with different symmetry label. For use in future analysis, we compute 
\begin{eqnarray}
\label{Eq_imbalance}
C_{\pm}&:=& + \frac{\gamma}{4}\mp \beta \frac{(1-\gamma)}{8}, \\ \nonumber
C_{2}&:=&  \beta \frac{(1-\gamma)}{4}= C_{-}-C_{+}, \\\nonumber
\end{eqnarray}
where $C_{\pm}$ is the imbalance of population between the $|A, m\rangle$ states and $|E_{\pm}, m'\rangle$ states and $C_{2}$ is the imbalance of population between $|E_{+} , m'\rangle$ states and $|E_{-} , m'\rangle$ states for all possible values of $m$ and $m'$ and at $t=0$.\\

 For the sake of abbreviation in notation, the population of the $x^{\text{th}}$ energy level at time $t$, is denoted by $[x]_{t}$ instead of $p_{x}(t)$, and the change of population of the $x^{\text{th}}$ level during the interval $t_{n-1}$ and $t_{n}$, is denoted by $\Delta [x]_{n}$ instead of $\Delta p_{x}(n)$. Thus, $[x]_{t_{n}} = [x]_{t_{n-1}} + \Delta[x]_{n}$. Additionally, we denote $J^{\lambda}_{2}(\omega_{s}+\omega_{I})$ with $ J^{\lambda}_{2}$ and $J^{\lambda}_{0}(\omega_{s}-\omega_{I})$ with $J^{\lambda}_{0}$.\\
 
 Considering an unpolarized test spin, $\alpha=0$, the solutions to the rate equations for short evolution time, $t=\delta t$, are
\begin{eqnarray}
\label{Eq_Sol_rate_equation1}
&\Delta [A, \frac{3}{2}, \uparrow]_{1} &=-\delta t \ \left(C_{+} \  J_{2}^{E_{-}} + C_{-}\  J_{2}^{E_{+}}\right),\\  \nonumber
 & \Delta[A, \frac{3}{2}, \downarrow]_{1} &=-\frac{1}{6}\delta t \ \left(C_{+} \  J_{0}^{E_{-}} + C_{-}\  J_{0}^{E_{+}}\right),\\  \nonumber \\ \nonumber \\ \nonumber
&\Delta[A, -\frac{3}{2}, \uparrow]_{1} &=-\frac{1}{6}\delta t \ \left(C_{+} \  J_{0}^{E_{+}} + C_{-}\  J_{0}^{E_{-}}\right),\\  \nonumber
&\Delta[A, -\frac{3}{2}, \downarrow]_{1} &=-\delta t \ \left(C_{+} \  J_{2}^{E_{+}} + C_{-}\  J_{2}^{E_{-}}\right),\\  \nonumber \\ \nonumber \\ \nonumber
&\Delta[E_{\pm}, \frac{1}{2}, \uparrow]_{1}&= \delta t \ \left( \frac{1}{3}\left( C_{\pm}  \ J_{2}^{E_{\pm}}  \mp  C_{2}\ J_{2}^{E_{\mp}}\right) + \frac{1}{ 6}C_{\pm} \ J_{0}^{E_{\mp}}\right),\\ \nonumber
&\Delta[E_{\pm}, \frac{1}{2}, \downarrow]_{1}&= \delta t \ \left( \frac{1}{3 \times 6}\left( C_{\pm}  \ J_{0}^{E_{\pm}}  \mp  C_{2}\ J_{0}^{E_{\mp}}\right) + C_{\pm} \ J_{2}^{E_{\mp}}\right),\\  \nonumber \\ \nonumber \\ \nonumber
&\Delta[E_{\pm}, -\frac{1}{2}, \uparrow]_{1}&= \delta t \ \left( \frac{1}{3 \times 6}\left(C_{\pm}  \ J_{0}^{E_{\mp}}  \mp  C_{2}\ J_{0}^{E_{\pm}}\right) + C_{\pm} \ J_{2}^{E_{\pm}}\right),\\ \nonumber
&\Delta[E_{\pm}, -\frac{1}{2}, \downarrow]_{1}&= \delta t \ \left( \frac{1}{3}\left( C_{\pm}  \ J_{2}^{E_{\mp}}  \mp  C_{2}\ J_{2}^{E_{\pm}}\right) + \frac{1}{ 6}C_{\pm} \ J_{0}^{E_{\pm}}\right).\\ \nonumber
 \end{eqnarray}
Because all energy levels in $A$ subspace are equally populated at $t=0$, in the absence of $DQ^{A}$ and $ZQ^{A}$ terms, we obtain $\Delta [ A, \pm \frac{1}{2},\uparrow]_{1}= \frac{1}{3} \Delta [ A, \pm \frac{3}{2} ,\uparrow]_{1} $ and $\Delta[ A, \pm \frac{1}{2} , \downarrow]_{1}= \frac{1}{3} \Delta[ A, \pm \frac{3}{2} , \downarrow]_{1}$. The expressions in Eq.\ref{Eq_Sol_rate_equation1}, may appear very complicated and it may sound difficult to get an insight about the relaxation. But, if we pay attention to the symmetry, there is a delicate and simple relation between the population of different energy levels. First of all, for all values of $(s, m)$, the change in the population of $ |s, m, \uparrow \rangle$ level is the same as that of $ |s,  m, \downarrow \rangle$ with just the difference of replacing $J^{\lambda}_{2} \leftrightarrow J^{\lambda}_{0}/6$. Second, the change of population in each level $|s, m, \uparrow \rangle$ is the same as that of $|s, -m, \uparrow \rangle$ with just the difference of replacing $J^{\lambda}_{2} \leftrightarrow J^{-\lambda}_{0}/6$. It will be shown that the anti-phase feature of the NMR peaks observed by others arises from these two properties, which are also visually captured from Fig.\ref{fig_non_zero_transitions}.\\

  For the $m=\pm \frac{1}{2}$ subspace, after summing over the symmetry labels and doing some algebra, we obtain
\begin{eqnarray}
\sum\limits_{s} \Delta[s,\pm\frac{1}{2} ,\uparrow]_{1} &=& - \Delta [A, \pm \frac{3}{2}, \downarrow]_{1},\\ \nonumber
\sum\limits_{s} \Delta[s,\pm\frac{1}{2} ,\downarrow]_{1} &=& - \Delta[A, \pm \frac{3}{2}, \uparrow]_{1},
\end{eqnarray}
where $s\in \{A, E_{\pm}\}$. We see in the following that these very neat relations between different energy levels enable us to anticipate the NMR signal analytically \\

To be more specific, we start from the NMR signal on the test spin channel. The scalar coupling $2\pi \ J_{1}\  \vec{\mathbf{S}}.\vec{I}$ that was neglected so far, shifts the frequency of the test spin condition on the total spin magnetization of protons. Therefore, it is expected to observe 4 distinguishable peaks on the $I$ channel corresponding to $m=\pm \frac{3}{2}$ and $m=\pm \frac{1}{2}$. We define a set of operators that project the collective spins into these magnetization subspaces with
\begin{eqnarray}
\Pi^{\pm \frac{3}{2}} &=& | A, \pm \frac{3}{2} \rangle \langle A, \pm \frac{3}{2}|, \\ \nonumber
\Pi^{\pm \frac{1}{2}} &=& \sum\limits_{s}  | s, \pm \frac{1}{2} \rangle \langle s, \pm \frac{1}{2}|.
\end{eqnarray}
 For a short evolution time, the expected NMR peaks at the test spin channel are
\begin{eqnarray}\nonumber
\label{Eq_NMR_peaks}
\langle \Pi^{\pm \frac{3}{2}} \otimes I_{z} \rangle\vert_{\alpha=0}^{\delta t} &=& \frac{1}{2}\left([A,\pm\frac{3}{2} ,\uparrow]_{\delta t} - [A,\pm\frac{3}{2} ,\downarrow]_{\delta t} \right)\\ \nonumber 
&=& \frac{\delta t}{8}\ \Big[ \pm \gamma \ \left(\Gamma^{E_{-}} +\Gamma^{E_{+}}\right)\\ \nonumber
&&\hskip 0.6 cm- \beta \frac{(1-\gamma)}{2}\   \left(\Gamma^{E_{+}} - \Gamma^{E_{-}}\right) \Big], \\ \nonumber
\langle \Pi^{\pm \frac{1}{2}} \otimes I_{z} \rangle\vert_{\alpha=0}^{\delta t} &=&\frac{1}{2}\left(\sum\limits_{s} [s,\pm\frac{1}{2} ,\uparrow]_{\delta t} - [s,\pm\frac{1}{2} ,\downarrow]_{\delta t} \right)\\ \nonumber 
&=& \langle \Pi^{\pm \frac{3}{2}} \otimes I_{z} \rangle\vert_{\alpha=0}^{\delta t},
\end{eqnarray}
in which $\Gamma^{\lambda}:= J_{2}^{\lambda} - \frac{1}{6} J_{0}^{\lambda}$. To compute the above NMR signals, we used $[x]_{\delta t} = [x]_{0} + \Delta[x]_{1}$ and replaced the expressions from Eq.\ref{Eq_Sol_rate_equation1} into it. Therefore, on the carbon channel, both $\mathbf{S}_{z} \otimes I_{z}$ and $\mathbb{1} \otimes I_{z}$ part of the density matrix become an NMR observable, leading to an anti-phase term that is proportional to $\gamma$ and an in-phase term that is proportional to $\beta$. Similarly, the anticipated NMR peaks on the proton channel are

\begin{eqnarray}\nonumber
\langle \mathbf{S}_{z} \otimes |\uparrow \rangle \langle \uparrow| \rangle\vert_{\alpha=0}^{\delta t} &=&\frac{3}{2}\ \left( [A, \frac{3}{2}, \uparrow]_{\delta t} - [A, -\frac{3}{2},\uparrow ]_{\delta t} \right)\\ \nonumber
&+& \frac{1}{2} \sum\limits_{s} \left([s, \frac{1}{2}, \uparrow]_{\delta t} - [s, -\frac{1}{2},\uparrow ]_{\delta t} \right) \\ \nonumber
&=&  \frac{\delta t}{8}\ \Big[  \gamma\ \left(\tilde{\Gamma}^{E_{+}} +\tilde{\Gamma}^{E_{-}}\right) \\ \nonumber
&&\hskip 0.6 cm+ \beta \frac{(1-\gamma)}{2}\ \left(\tilde{\Gamma}^{E_{+}} - \tilde{\Gamma}^{E_{-}}\right) \Big]\\ \nonumber
&=& - \langle \mathbf{S}_{z} \otimes |\downarrow \rangle \langle \downarrow| \rangle\vert_{\alpha=0}^{\delta t} 
\end{eqnarray}
in which $\tilde{\Gamma}^{\lambda}:= J_{2}^{\lambda} - \frac{1}{6} J_{0}^{-\lambda}$. Therefore, on the proton channel, $ \mathbf{S}_{z} \otimes I_{z}$ becomes an NMR observable, leading to two equal peaks with opposite phases.\\


It remains to solve the rate equations in case of $\alpha\neq 0$. When the test spin has some initial polarization, $\alpha\neq 0$, in Eq.\ref{Eq_imbalance} additional terms show up. Precisely, the imbalance of population difference between different energy levels is now replaced by
\begin{eqnarray}
\label{Eq_imbalance_p}
[A,m, \uparrow]_{0}-[E_{\pm},m',\downarrow]_{0}&=& C_{\pm} +  \left(\frac{\alpha}{4}  \pm \alpha \beta \frac{(1-\gamma)}{8}\right),\\ \nonumber
  [A, m, \downarrow]_{0}-[E_{\pm}, m',\uparrow]_{0}&=& C_{\pm} -  \left(\frac{\alpha}{4}  \pm \alpha \beta \frac{(1-\gamma)}{8}\right), \\ \nonumber
[E_{+},m', \uparrow]_{0}-[E_{-}, m',\downarrow]_{0} &=& C_{2} +   \alpha \frac{\ (1-\gamma)}{4},\\ \nonumber
[E_{+},m',\downarrow]_{0}-[E_{-},m',\uparrow]_{0} &=& C_{2} -   \alpha \frac{\ (1-\gamma)}{4},
\end{eqnarray}
$ \forall m\in \{\pm \frac{3}{2}, \pm\frac{1}{2}\} $ and $\forall m'\in \{\pm \frac{1}{2}\}$. By replacing these initial imbalance of population in the rate equation (Eq.\ref{Eq_rate_eq}) and after doing some tedious calculations, we obtain 

\begin{eqnarray}
\label{Eq_peaks}
&&\langle \Pi^{\pm \frac{3}{2}} \otimes I_{z} \rangle \vert_{\alpha\neq 0}^{\delta t} - \langle \Pi^{\pm \frac{3}{2}} \otimes I_{z} \rangle \vert_{\alpha= 0}^{\delta t} = \\ \nonumber
&& \frac{\delta t}{8}\left[  \alpha\ \left(\tilde{\Gamma}^{E_{+} }+ \tilde{\Gamma}^{E_{-}}\right) \pm \alpha \beta \frac{(1-\gamma)}{2} \ \left( \tilde{\Gamma}^{E_{+}} - \tilde{\Gamma}^{E_{-}}\right) \right] \\  \nonumber \\ \nonumber\\ \nonumber
 &&\langle \Pi^{\pm \frac{1}{2}} \otimes I_{z} \rangle \vert_{\alpha\neq 0}^{\delta t} - \langle \Pi^{\pm \frac{1}{2}} \otimes I_{z} \rangle \vert_{\alpha=0}^{\delta t} = \\ \nonumber
 && \frac{\delta t}{8} \left[  \alpha\ \left(\tilde{\Gamma}^{E_{+} }+ \tilde{\Gamma}^{E_{-}}\right) \pm \alpha \frac{(2\beta-1)}{3} \frac{(1-\gamma)}{2} \ \left( \tilde{\Gamma}^{E_{+}} - \tilde{\Gamma}^{E_{-}}\right) \right]
\end{eqnarray}

Interestingly, all terms that have $\beta$ dependency are proportional to $J^{E_{+}}(\omega) - J^{E_{-}}(\omega)$. This means that the polarization difference between the $E_{+}$ and the $E_{-}$ subspaces becomes observable if $J^{E_{+}}(\omega) \neq J^{E_{-}}(\omega)$. This also means that if one is able to create an imbalance of population between the $E_{+}$ and the $E_{-}$ subspaces, that wavefunction leads to an extra feature in the observed NMR spectra which reveals information about the symmetry of the spectral density of noise. \\

In case of $J^{E_{+}}(\omega)= J^{E_{-}}(\omega)$, Eq.\ref{Eq_peaks} reduces to
\begin{eqnarray}\nonumber
\langle \Pi^{\pm \frac{3}{2}} \otimes I_{z} \rangle \vert_{\alpha\neq 0}^{\delta t} &=& \frac{\delta t}{8}\ \left[ \pm \gamma \ \left(\Gamma^{E_{-}} +\Gamma^{E_{+}}\right) -   \alpha\ \left(\tilde{\Gamma}^{E_{+} }+ \tilde{\Gamma}^{E_{-}}\right)  \right] \\ \nonumber
\langle \Pi^{\pm \frac{1}{2}} \otimes I_{z} \rangle \vert_{\alpha\neq 0}^{\delta t} &=& \frac{\delta t}{8}\ \left[ \pm \gamma \ \left(\Gamma^{E_{-}} +\Gamma^{E_{+}}\right) -   \alpha\ \left(\tilde{\Gamma}^{E_{+} }+ \tilde{\Gamma}^{E_{-}}\right)  \right] \\ \nonumber
\end{eqnarray}
The anti-phase contribution in the NMR spectra is proportional to $\gamma$ which is the initial imbalance of population between the $A$ symmetry subspace and the $E$ symmetry subspace. The in-phase contribution is proportional to $\alpha$ which is the polarization of the test spin that is dipole coupled with the three protons.  Because we do not know the explicit value of different orders of the spectral density of noise, we cannot make any conclusion about the relative magnitude of these two contributions. Nevertheless, we can conclude that the peaks in pair of $(\frac{3}{2},\frac{1}{2})$ or $(-\frac{3}{2}, -\frac{1}{2})$ have the same phase and the same amplitude. Now, if $J^{E_{+}}(\omega)\neq J^{E_{-}}(\omega)$, even for the case of $\beta =0$, we can no longer make any judgment about the relative amplitude and phase of these 4 peaks. Because, according to Eq.\ref{Eq_peaks}, even when $\beta =0$ an additional anti-phase term survives for the $m=\pm \frac{1}{2}$ peaks. In Appendix.\ref{Apdx_Rate_Equation}, we compute the NMR peaks for one step further when $t=2 \delta t$ and conclude that for a longer evolution time, additional terms show up that diffrentiates the amplitude of the $m=\pm \frac{3}{2}$ from that of $m=\pm \frac{1}{2}$. Thus, 4 peaks with unequal amplitudes is expected. This argument holds true without any assumption about the magnitude of the anti-phase contribution relative to the in-phase contribution. Now, we assume those terms with a $\gamma$ factor are larger than the others and conclude that the $m$ peaks are overall anti-phase with the $-m$ peaks but with unequal amplitudes. A large $\gamma$ corresponds to the long lived state that was experimentally demonstrated in \cite{M13}, and surprisingly the above model, which considers only the $DQ^{E_{\pm}}$ and the $ZQ^{E_{\pm}}$ transitions of the DD coupling between the collective spin and an external spin, results in an analytic solution that predicts most features of the NMR peaks that have been experimentally observed.
\subsection{Relaxation Induced by Homonuclear Coupling} 
In case of internal dipolar coupling the analogy to Eq.\ref{Eq_master_dipolar2} is
 \begin{eqnarray}
 \label{Eq_master_dipolar3}
 \frac{\partial \tilde{\rho}}{\partial t} &=& \sum\limits_{\lambda} \  \mathbb{ZQ}^{\lambda}[\tilde{\rho}] +  \mathbb{DQ}^{\lambda}[\tilde{\rho}] +  \mathbb{SQ}^{\lambda}[\tilde{\rho}] \\ \nonumber
\mathbb{ZQ}^{\lambda}[.]&:=& \frac{1}{6}\ g^{\lambda}_{0}(0)\  \left( \hat{D}[ \mathbb{T}_{0}^{\lambda}][.] + \hat{D}[ \mathbb{T}_{0}^{-\lambda}][.] \right) \\ \nonumber
  \mathbb{DQ}^{\lambda}[.]&:=&  g^{\lambda}_{2}(2\ \omega_{s})\  \left( \hat{D}[\mathbb{T}_{2}^{\lambda}][.] + \hat{D}[\mathbb{T}_{-2}^{-\lambda}][.] \right) \\ \nonumber
  \mathbb{SQ}^{\lambda}[.]&:=& g^{\lambda}_{1}( \omega_{s}) \ \left( \hat{D}[ \mathbb{T}_{1}^{\lambda}][.] + \hat{D}[ \mathbb{T}_{-1}^{-\lambda}][.]  \right) \\ \nonumber.
 \end{eqnarray}
 where 
 \begin{equation}
 g^{\lambda}_{q}( \omega):=  \int\limits_{-\infty}^{\infty} \overline{G_{q}^{\lambda}(t) \ G_{q}^{\lambda*}(t+ \tau)} \ e^{-i \omega\tau} \ d\tau
 \end{equation}
 is the spectral density of noise due to the homonuclear dipolar Hamiltonian. We simply denote $g^{\lambda}_{q}( q\ \omega_{s})$ by $g^{\lambda}_{q}$. \\
 
 We solve the master equation using the same approach we used in the previous section for hetronuclear coupling. In order to calculate the change of population of each energy level during the short evolution time $\delta t$, we need to know which transitions are allowed (the non-zero matrix elements of $\mathbb{T}^{\lambda}_{q}$ as shown in Fig. \ref{fig_Homo_Transitions}) and the initial imbalance of population between the allowed transitions. Thus, for each energy level $ x \in \{|s, m\rangle\}$, we should calculate $\Delta [x]_{ 1} = \sum\limits_{y} \ ([y]_{0} - [x]_{0}) \times W_{xy}$ in which $W_{xy}$ are the transition rates and $ [x]_{0}$ is the population of the $x^{\text{the}}$ level at $t=0$.\\
 
  For the initial state of interest (Eq.\ref{Eq_Initial_state}), at $t=0$, all the energy levels $| s, \frac{1}{2}\rangle$ have the same population as $| s,- \frac{1}{2}\rangle$ energy levels. On the other hand, the transition rates induced by $\mathbb{ZQ}^{\lambda}$  terms of the internal dipolar coupling is identical for both $| s, \frac{1}{2}\rangle$ states and $| s, -\frac{1}{2}\rangle$. As a result, the change of population of $|s, m\rangle $ is equal to that of f $|s, -m\rangle $, leading to a zero polarization in the Zeeman basis. Therefore, even though there is an exchange of population between different symmetry subspaces, as shown in the middle section of Fig.\ref{fig_Homo_Transitions}, the net contribution of flip-flop interaction among protons to the NMR spectra of a protected state is zero. \\
  
 
Considering the above point, the solution of the master equation in Eq.\ref{Eq_master_dipolar3} for short evolution time, $\delta t$, is

\begin{eqnarray}
\label{Eq_Sol_rate_equation2}
\Delta [A,  \pm \frac{3}{2}]_{1} = &- &  \frac{\delta t}{4} \ \left( C_{+}\  g_{1}^{E_{\mp}} + C_{-} \ g_{1}^{E_{\pm}}\right)\\  \nonumber
&-& \delta \ t  \left( C_{+}\  g_{2}^{E_{\mp}} + C_{-} \ g_{2}^{E_{\pm}}\right)\\ \nonumber
\Delta [A,  \pm \frac{1}{2}]_{1} = &- &  \frac{3\ \delta t}{4} \ \left( C_{+}\  g_{1}^{E_{\mp}} + C_{-} \ g_{1}^{E_{\pm}}\right) \\ \nonumber
\Delta [E_{+},  \pm \frac{1}{2}]_{1} = &+ &\  \frac{\delta t}{4} \  C_{+} \left( g_{1}^{E_{\mp}} + 3\ g_{1}^{E_{\pm}} \right)\\ \nonumber
&+& \delta \ t\ C_{+} \ g_{2}^{E_{\pm}} \\ \nonumber
\Delta [E_{-},  \pm \frac{1}{2}]_{1} = &+ &\  \frac{\delta t}{4} \  C_{-} \left( g_{1}^{E_{\pm}} + 3\ g_{1}^{E_{\mp}}\right) \\ \nonumber
&+& \delta \ t\ C_{-} \ g_{2}^{E_{\mp}} \\ \nonumber
 \end{eqnarray}
Given the above relations and $[x]_{\delta t}= [x]_{0} + \Delta [x]_{1}$ for all $x \in \{ |s, m\rangle \}$, the expected NMR spectrum of protons for short evolution time is
\begin{eqnarray}\nonumber
\langle \mathbf{S}_{z}  \rangle_{\delta t} &=&\frac{3}{2}\ \left( [A, \frac{3}{2}]_{\delta t} - [A, -\frac{3}{2} ]_{\delta t} \right)\\ \nonumber
&+& \frac{1}{2} \sum\limits_{s} \left([s, \frac{1}{2}]_{\delta t} - [s, -\frac{1}{2} ]_{\delta t} \right) \\ \nonumber
&=& \delta t \ (C_{+} - C_{-}) \ \left( (g_{1}^{E_{+}}- g_{1}^{E_{-}}) + 2\ (g_{2}^{E_{+}}- g_{2}^{E_{-}})  \right)
\end{eqnarray}

 We repeat the above calculation for the second time step, $t= 2\ \delta t$, using the method explained in Appendix. \ref{Apdx_Rate_Equation}  and concluded that the signal is proportional to $(C_{+} - C_{-}) \ ( g_{q} ^{E_{+}} + g_{q}^{E_{-}}) $. \textit{This means that if the $E_{+}$ and the $E_{-}$ subspaces are equally populated at $t=0$, the higher order contribution of the homonuclear dipolar relaxation to the NMR spectra is also zero}. We justify this counter-intuitive result by comparing the spin operators of the hetronculear  coupling to that of the homonuclear couplings. The $ZQ^{\lambda}$ and the $DQ^{\lambda}$ transitions of the external dipolar interaction occur when the protons and of methyl group exchange energy with an external spin. This \textit{effectively} appears as a single spin flip (or a $\mathbf{S}_{\pm}^{\lambda}$ operator) on the space of protons. In that sense the $T_{+}^{\lambda}$ of the homonuclear coupling is similar to the $\mathbf{S}_{\pm}^{\lambda}$ of the hetronuclear coupling. But, there are two important points that differentiates the two relaxation processes. The first point is that the $T_{+}^{\lambda}$ is block-diagonal in the $E$ subspace, meaning that there is no exchange of population between the $E_{+}$ and the $E_{-}$ subspaces as opposed to the $\mathbf{S}_{+}^{\lambda}$ operator. As a result, when these two spaces are equally populated at $t=0$, after an evolution time $\delta t$, the net polarization in the Zeeman basis remains zero, because, both of these subspaces exchange population with the $A$ subspace with the same rate but in opposite direction. Thus, the homonuclear coupling \textit{passively} breaks the symmetry of an initially prepared protected state meaning that even thought it induces non-zero transition between different symmetry subspaces, its effective contribution to the NMR observation of LLS is yet zero. The second point is that, in case of hetronuclear interaction, when a $|s, m, \uparrow \rangle$ level experiences a $DQ$ transition with the rate $J_{2}^{\lambda}$ for instance, its counterpart $|s, -m, \uparrow \rangle$ experiences a $ZQ$ transition with the rate $\frac{1}{6}J_{0}^{\lambda}$. Therefore, the differences between the two rates, $(J^{\lambda}_{2} - \frac{1}{6} \ J^{\lambda}_{0})$ appears as a non-zero factor in the NMR signal (Eq.\ref{Eq_NMR_peaks}). This holds true even if the $E_{+}$ and the $E_{-}$ subspaces have identical initial population. Thus, the hetrouclear coupling \textit{actively} breaks the symmetry of an initially prepared protected state because it leads to a non-zero signal.  
%
\section{Conclusion}
\label{Sec_Conclude}
We analyzed the contribution of both the internal and the external dipolar interaction of methyl groups in NMR observation of long lived states that are initially prepared as symmetry polarized states. The symmetry properties of spin operators and the density matrix allowed us to first show why these states exhibit long relaxation times and second to solve the master equation analytically and obtain the NMR spectrum. Our theoretical result is in agreement with the reported experimental observations. We find that even though both the hetronuclear and the homonuclear dipolar couplings break the symmetry of LLS, only the former leads to a relaxation pathway with non-zero polarization trasnfer from symmetry order to Zeeman order. This approach might find application in other areas in NMR where the underlying physics of the relaxation mechanisms are of interest.\\

The research results communicated here would not be possible without the significant contributions of the Canada First Research Excellence Fund, Industry Canada, the Natural Sciences and Engineering Research Council of Canada (NSERC RGPIN-418579), Canada Excellence Research Chairs Program (CERC 215284), the Canadian Institute for Advanced Research, the Quantum NanoFab and the Province of Ontario.
 
\appendix
\section{}
\label{Apdx_Rate_Equation}
In the main text, we solved the rate equations for short evolution time, $t=\delta t$. To compute the change of population of the $x^{\text{th}}$ energy level during $t_{0}$ and  $t_{0}+ \delta t$, it is required to know the population difference between the $x^{\text{th}}$ level and all other energy levels $y$ at time $t_{0}$ which are allowed due to the $DQ^{\lambda}$ and $ZQ^{\lambda}$ terms of DD coupling. For the first step, when $t_{0}=0$, the calculation of population difference between allowed transitions yields Eq.\ref{Eq_imbalance} when the test spin is unpolarized, and results in Eq.\ref{Eq_imbalance_p} when the test spin has a polarization $\alpha \neq 0$. In this appendix, we move one step forward and compute the change of population of the $x^{\text{th}}$ level during $t=\delta t$ and $t=2\delta t$. This requires the calculation of population difference between the allowed transitions, $x$ and $y$, at an earlier time, $t_{0}=\delta t$. This is computed by 

\begin{equation}
 [x]_{\delta t}-[y]_{\delta t} = [x]_{0}-[y]_{0} + \Delta[x]_{1}- \Delta[y]_{1}.
\end{equation}

The first two terms, $[x]_{0}-[y]_{0} $ are already calculated in Eq.\ref{Eq_imbalance_p}. To compute the second two terms, we assume $\alpha=0$, for simplicity, and use the result in Eq.\ref{Eq_Sol_rate_equation2}. 
 We start from $m= \pm \frac{3}{2}$ and obtain
\begin{eqnarray}
\label{Eq_pop_diff_dt}
\Delta[A, \frac{3}{2}, \uparrow]_{1}-\Delta[E_{\pm}, \frac{1}{2}, \downarrow]_{1} &=& - \delta t\ ( a_{0}  + a_{\pm}) \\ \nonumber
\Delta[A, \frac{3}{2}, \downarrow]_{1}-\Delta[E_{\pm}, \frac{1}{2}, \uparrow]_{1} &=& - \delta t\ ( b_{0}  + b_{\mp}) \\ \nonumber
\Delta[A,- \frac{3}{2}, \downarrow]_{1}-\Delta[E_{\pm}, -\frac{1}{2}, \uparrow]_{1} &=& - \delta t\ ( a_{0}  + a_{\mp}) ^{\beta \rightarrow -\beta}\\ \nonumber
\Delta[A,- \frac{3}{2}, \uparrow]_{1}-\Delta[E_{\pm}, -\frac{1}{2}, \downarrow]_{1} &=& - \delta t\ ( b_{0}  + b_{\mp}) ^{\beta \rightarrow -\beta}
\end{eqnarray}
in which $a_{0}$, $b_{0}$, $a_{\pm}$ $b_{\pm}$ are constants and the subscript $\beta \rightarrow -\beta$ means that in the definition of these constants, $\beta$ is replaced with $-\beta$. Those constants are explicitly
\begin{eqnarray}\nonumber
a_{0} &=& \frac{\gamma}{4}\ ( J^{E_{+}}_{2}+J^{E_{-}}_{2} + \frac{J^{E_{+}}_{0}+J^{E_{-}}_{0}}{18}) + \beta \frac{(1-\gamma)}{8} \ (J^{E_{+}}_{2}-J^{E_{-}}_{2}) \\ \nonumber
b_{0} &=& \frac{\gamma}{4}\ ( \frac{J^{E_{+}}_{0}+J^{E_{-}}_{0}}{6} + \frac{J^{E_{+}}_{2}+J^{E_{-}}_{2}}{3}) + \beta \frac{(1-\gamma)}{8}\ (J^{E_{+}}_{2}-J^{E_{-}}_{2}) \\ \nonumber
a_{\pm} &=& \frac{\gamma}{4}\ ( J^{E_{\pm}}_{2}- \frac{J^{E_{\pm}}_{0}}{18}) \pm \beta \frac{(1-\gamma)}{8}\  (\frac{J^{E_{\pm}}_{0}}{9}+ \frac{J^{E_{\mp}}_{0}}{18}) \\ \nonumber
b_{\pm} &=& \frac{\gamma}{4}\ ( \frac{J^{E_{\pm}}_{0}}{6}- \frac{J^{E_{\pm}}_{2}}{3}) \pm \beta \frac{(1-\gamma)}{8}\  (\frac{2J^{E_{\pm}}_{2}}{3}+ \frac{J^{E_{\mp}}_{2}}{3}) 
\end{eqnarray} 
The expressions in Eq.\ref{Eq_pop_diff_dt} gives the population difference between allowed transitions at time $t_{0}= \delta t$. Thus, we substitute  Eq.\ref{Eq_pop_diff_dt} into the rate equation, Eq.\ref{Eq_rate_eq}, in order to calculate $\Delta[x]_{2}$ and the result is
\begin{widetext}
\begin{eqnarray}
\label{Eq_up_3_2}
\Delta [A, \frac{3}{2}, \uparrow]_{2}&=& \Delta[A, \frac{3}{2}, \uparrow]_{1} + \delta t^{2} \ ( (a_{0} + a_{+}) \ J^{E_{+}}_{2} +(a_{0} + a_{-}) \ J^{E_{-}}_{2} ) \\ \nonumber
\Delta [A, \frac{3}{2}, \downarrow]_{2}&=& \Delta[A, \frac{3}{2}, \downarrow]_{1} + \frac{\delta t^{2}}{6} \ ( (b_{0} + b_{+}) \ J^{E_{+}}_{0} +(b_{0} + b_{-}) \ J^{E_{-}}_{0} ) \\ \nonumber \\ \nonumber
\Delta [A, -\frac{3}{2}, \uparrow]_{2}&=& \Delta[A, -\frac{3}{2}, \uparrow]_{1} + \frac{\delta t^{2}}{6} \ ( (b_{0} + b_{-}) \ J^{E_{-}}_{0} +(b_{0} + b_{+}) \ J^{E_{+}}_{0} ) ^{\beta \rightarrow -\beta}\\ \nonumber
\Delta [A, -\frac{3}{2}, \downarrow]_{2}&=& \Delta[A, -\frac{3}{2}, \downarrow]_{1} + \delta t^{2} \ ( (a_{0} + a_{-}) \ J^{E_{-}}_{2} +(a_{0} + a_{+}) \ J^{E_{+}}_{2} ) ^{\beta \rightarrow -\beta}
\end{eqnarray}
\end{widetext}
Then, to compute the NMR signal on the test spin channel condition on the collective spin magnetization $m=\pm\frac{3}{2}$, we subtract the above expressions from each other and in the calculation, we encounter terms like this
$$ (a_{0}+ a_{+}) J^{E_{+}}_{2} - (b_{0} + b_{+}) \frac{1}{6}J_{0}^{E_{+}} = \frac{\gamma}{4}\ A + \beta \frac{1- \gamma}{8}\ B$$
which is written in such a way that $A$ contains all terms with $\frac{\gamma}{4}$ dependency and $B$ contains all terms with $\beta \frac{1- \gamma}{8}$. Intuitively, if $\beta=0$, the NMR signal from $\pm m$ have the same amplitude but with opposite sign. This is because the counter part of $|A\ ,\  +\frac{3}{2}, \uparrow\rangle$ is $|A\ ,\ -\frac{3}{2}, \downarrow\rangle$ where both are involved in $DQ$ transition and the counterpart of $|A, +\frac{3}{2}, \downarrow\rangle$ is $|A, -\frac{3}{2}, \uparrow\rangle$ where both are involved in $ZQ$ transition. Thus, the contribution from the $\frac{\gamma}{4}$ dependent terms results in an anti-phase signal from $m=\pm \frac{3}{2}$ peaks. In contrast, the contribution from the $\beta$ dependent terms results in an in-phase signal from $m=\pm \frac{3}{2}$ peaks, because, $m=-\frac{3}{2}$ picks an additional minus when $\beta \rightarrow -\beta$. Therefore, given the expressions in Eq.\ref{Eq_up_3_2} and the first order calculation that was given in the main text in Eq.\ref{Eq_NMR_peaks}, the second order anticipated NMR signal for evolution time, $t=2\ \delta t$ is
\begin{eqnarray}\nonumber
\langle \Pi^{\pm \frac{3}{2}} \otimes I_{z} \rangle_{2\delta t} &=& \langle \Pi^{\pm  \frac{3}{2}} \otimes I_{z} \rangle_{\delta t} +  \frac{\delta t^2}{8}\ \{  \pm \gamma \ A + \beta \frac{(1-\gamma)}{2}\  B \}
\end{eqnarray} 
The same logic holds for $m=\pm\frac{1}{2}$ NMR peaks and similar feature is obtained,
\begin{eqnarray}\nonumber
\langle \Pi^{\pm \frac{1}{2}} \otimes I_{z} \rangle_{2\delta t} &=& \langle \Pi^{\pm  \frac{1}{2}} \otimes I_{z} \rangle_{\delta t} +  \frac{\delta t^2}{8}\ \{  \pm \gamma \ A' + \beta \frac{(1-\gamma)}{2}\  B' \}
\end{eqnarray}
It is important to note that in contrast to the calculated NMR signals for $t=\delta t$, here at $t=2 \delta t$, the NMR peaks of $m=\pm\frac{1}{2}$ do not have the same amplitude as that of $m=\pm\frac{3}{2}$. This has roots in the fact that at $t=\delta t$, all energy levels within each symmetry subspace are not equally populated any more, leading to  $A\neq A'$ and $B\neq B'$.

To conclude, the solution of the rate equations for the second step, $t= 2\delta t$, has similar features to that of Eq.\ref{Eq_Sol_rate_equation1}, and therefore, all arguments in the main text regarding the phase and amplitude of the NMR peaks, which were concluded for short evolution $t= \delta t$, can be generalize to any time evolution by inductive reasoning. 

\
\bibliographystyle{unsrt}
\renewcommand\bibname{References}
\bibliography{Bibliography}
%
%

\end{document}